\documentclass[12pt]{iopart}

\usepackage{iopams}  

\usepackage{url}

\usepackage{bm}
\usepackage{amssymb}
\usepackage{graphicx}
\usepackage{float}
\usepackage{floatflt}

\begin{document}

\title{Life, the universe, and everything -- 42 fundamental questions}

\author{Roland E. Allen}
\address{Department of Physics and Astronomy, Texas A\&M University \\ College Station, Texas 77843, USA}
\author{Suzy Lidstr\"{o}m} 
\address{Department of Physics and Astronomy, Uppsala University \\ 
SE-75120 Uppsala, Sweden}
\smallskip 
\address{Physica Scripta, Royal Swedish Academy of Sciences \\ SE-104 05 Stockholm, Sweden}

\begin{abstract}
In \textit{The Hitchhiker's Guide to the Galaxy}, by Douglas Adams, the Answer to the Ultimate Question of Life, the Universe, and Everything is found to be 42 -- but the meaning of this is left open to interpretation. We take it to mean that there are 42 fundamental questions which must be answered on the road to full enlightenment, and we attempt a first draft (or personal selection) of these ultimate questions, on topics ranging from the cosmological constant and origin of the universe to the origin of life and consciousness.
\end{abstract}

\maketitle

\section{\label{sec:sec1} Motivation for this article}

Each dramatic new discovery in fundamental physics (including astrophysics) has revealed new features of our universe and new mysteries. The most challenging current issues range from the cosmological constant problem to the origin of spacetime and quantum fields, and include emergent phenomena such as life and consciousness. 

This article is addressed to several different groups of readers. We first wish to demonstrate to young scientists that they have at least as much opportunity to make major contributions to human understanding as they would have had in any previous century. A second goal is to introduce general readers to the most fundamental problems in science, with a presentation that may be more satisfying than either more technical papers or less informative popular accounts. And finally, we invite experts in the scientific community to write reviews of topics like those considered here, to form a series of invited articles similar in format to those which have recently appeared in this journal.  

This is an optimistic article, written largely to counteract a widespread perception that fundamental science has lost its excitement, as reflected in the pessimistic tone of titles like \textit{The End of Science}~\cite{endsci} or \textit{The End of Physics}~\cite{endphys}. Our main point is this: Never in history has there been a larger set of truly fundamental questions, and the answers to many seem tantalizingly close.

It should, of course, be emphasized that the path to revolutionary discoveries is rarely direct and intentional. Instead, it usually involves patient investigations that are motivated by curiosity and carried out with care and hard work. 

\begin{figure}[htbp]
\centering
\includegraphics[bb=0 0 360 600, width=3in]{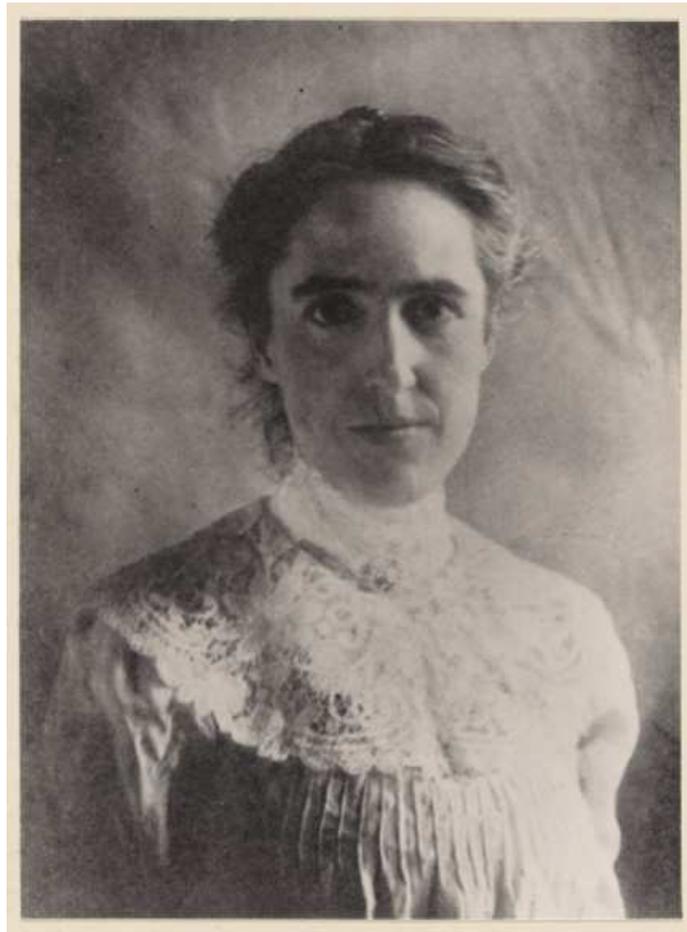}
 \caption{Henrietta Leavitt discovered the relation between the intrinsic brightness of a variable star and its period of oscillation, thereby giving astronomers their primary tool for measuring distances. 
Credit: Schlesinger Library, Radcliffe Institute, Harvard University.
\label{leavitt}}
 \end{figure}

One example -- or role model -- is Henrietta Leavitt, shown in Fig.~\ref{leavitt}, who studied the images of 1,777 variable stars recorded on photographic plates. She discovered that the time period over which the brightness of the star varies is an accurate measure of the intrinsic brightness of the star, with a brighter star oscillating more slowly.  When this intrinsic brightness is compared with the apparent brightness as seen from Earth, one can calculate the distance of the star. 
Her discovery of this phenomenon led to many other major discoveries in astronomy by Edwin Hubble and others, including the facts that there are many galaxies beyond our own Milky Way and that the universe is expanding. It is not an overstatement to say that Cepheid variable stars are the first and most widely used ``standard candles'' of astronomy, and that Henrietta Leavitt provided the key to determining the true size and behavior of the entire cosmos.
\begin{figure}[htbp]
\centering
 \includegraphics[bb=0 0 400 390, width=4in]{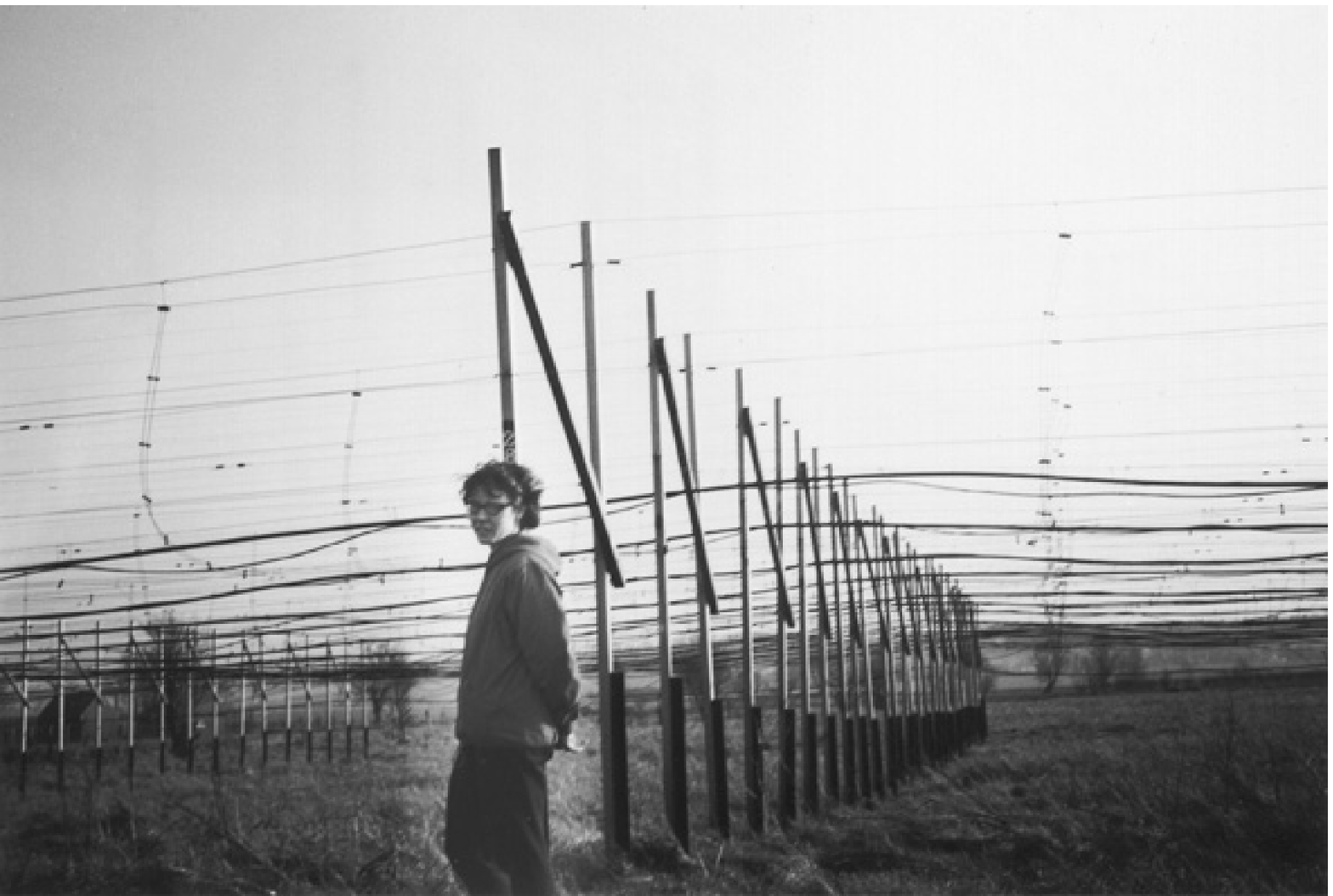}
 \caption{Jocelyn Bell Burnell discovered pulsars (rapidly spinning neutron stars), one of the greatest breakthroughs in the history of astronomy. She was placed in charge of operating a radio telescope which she had helped to construct -- and with which she is shown here. She then detected the unusual pulsar signal while analyzing the data. Her advisor, Antony Hewish, also played an important role in the discovery.  Credit: Jocelyn Bell Burnell.}
\label{Bell-telescope}
 \end{figure}

Figure~\ref{Bell-telescope} shows another central figure in the history of modern astronomy -- Jocelyn Bell Burnell, with the four-acre radio telescope which she used in the meticulous observations and data analysis that led to her discovery of pulsars (in collaboration with her advisor, Antony Hewish). Neutron stars had been proposed in 1934 by Walter Baade and Fritz Zwicky (see Fig.~\ref{Zwicky}), but the discovery of their actual existence led to an explosion of theoretical and observational studies which began in the late 1960s and continues until today. For example, the first demonstration of gravitational waves was provided by the orbital motion of a binary pulsar system. Once again, a great discovery came only after hard work, including even physical work in the field. She is quoted as saying ``By the end of my PhD I could swing a sledgehammer.''

If one turns from astronomy and cosmology to biology, one finds a comparable role model in Rosalind Franklin. Her experimental findings -- using X-ray diffraction -- provided the foundation for understanding the structure and role of DNA, which is commonly regarded as the basis for all life on Earth. 

Continuing with this same theme -- how patient experimentation can lead to revolutionary discoveries -- the perfect and most obvious example in physics and chemistry is Marie Curie, the only  Nobel Laureate in both fields. She played a major role in discovering (and naming) the phenomenon of radioactivity. While raising two daughters, she undertook the extremely arduous task of isolating radium, with tons of pitchblende yielding one-tenth of a gram of radium chloride. She later wrote ``Sometimes I had to spend a whole day stirring a boiling mass with a heavy iron rod nearly as big as myself. I would be broken with fatigue at day's end.''~\cite{curie-quote} 

Patience and hard work are also required in mathematical investigations. Here it is appropriate to mention Emmy Noether, who developed the profound insight that is now called Noether's theorem: Symmetries lead to conservation laws, so that the invariance of the laws of Nature under shifts in time, displacements in spatial position, rotations, and the gauge transformations of electromagnetism leads to conservation of energy, momentum, angular momentum, and electric charge. Her work illustrates a principle with many other examples: Mathematicians who are motivated by impractical considerations -- such as beauty or the appeal of an intellectual challenge -- often make contributions of enormous practical importance.

In the next sections we will list many fundamental mysteries or problems, and the above examples are meant to illustrate how such problems are usually solved in practice -- by hard work and the patient accumulation of understanding. Einstein's $E=mc^2$ did not suddenly pop into his head; it instead resulted from about ten years of thought on issues related to motion and electromagnetism, with a knowledge of the relevant experiments. His theory of gravity (or general theory of relativity) was again embedded in the real world of experiment and observation, including the fact that precise observations and equally precise calculations left $43$ of $574$ seconds of arc per century unaccounted for in the precession of the orbit of Mercury around the Sun. It was this legacy of earlier work that allowed Einstein to make a quantitative comparison with his calculations, which themselves followed a second ten-year intellectual Odyssey.

\section{\label{sec:sec2}Gravitational and cosmological mysteries}

Many of the following topics have extremely long histories (a century in the case of quantum gravity), with hundreds of important papers. A complete list of references would in fact contain far more than a thousand citations, if it were to properly credit all the original discoveries and ideas. Since such a list is prohibitive, we instead attempt to list a few representative later papers, reviews, and books which themselves cite the original work. There is heavy reliance on the reviews in a recent Nobel Symposium, because these papers are high-quality, timely, and free. In the present paper the description of both the science and the history has been, of necessity, very much simplified.

There are many previous lists or discussions of fundamental questions, some of which are excellent~\cite{ginzburg,duff-list,carroll} but have a different emphasis. (Refs.~\cite{ginzburg} and \cite{duff-list} are somewhat analogous to Hilbert's famous list of mathematical problems~\cite{hilbert}.) For example, we consider a number of problems which are so fundamental that they extend outside the normal range of current research. However, we limit attention to issues that are genuinely scientific in the broadest sense. For example, it has been suggested that we should include ``Does God exist?'', which we would rephrase as ``How should one define God to be able to truthfully say that God exists?'' (with the limiting case being ``God $=$ Nature''). But we omit questions like this from consideration. It has also been suggested that the question of greatest interest to most human beings is ``How should we live?'', which we would rephrase as ``Is there a meaningful ethical system which is as well defined as mathematics?'' (in the sense that one could formulate reasonable ethical axioms that unambiguously specify correct behavior). But we again omit this question as one that is not relevant to understanding Nature.

\subsection{\label{subsec:subsec1}The cosmological constant problem}

\begin{figure}[htbp]
\centering
\includegraphics[bb=0 0 360 1000, width=1.7in]{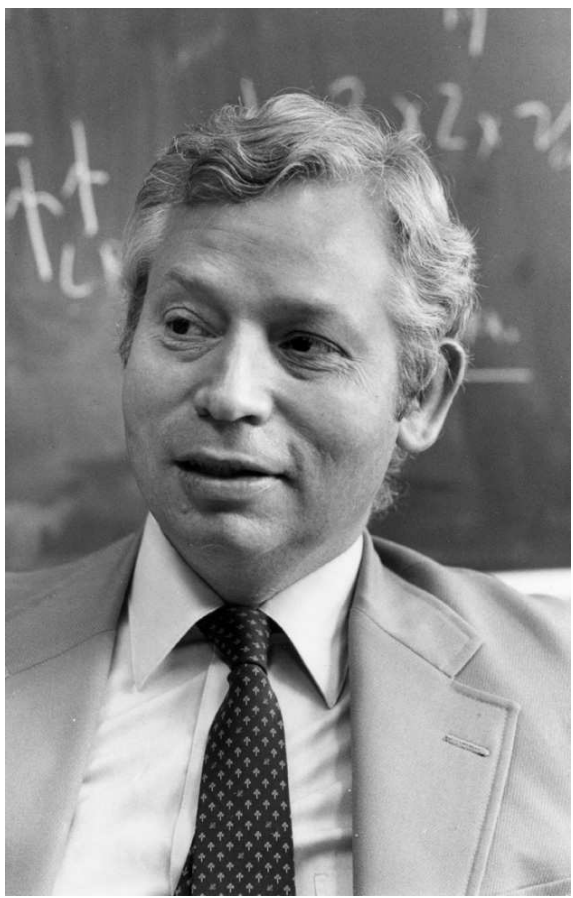}
 \caption{Steven Weinberg is a principal architect of the Standard Model of particle physics. He has also been active in gravitation, cosmology, and other areas of theoretical physics. 
Credit: AIP Emilio Segre Visual Archives.
\label{weinberg}}
 \end{figure}
The cosmological constant problem was thoroughly discussed in a 1989 paper by Steven Weinberg~\cite{weinberg-1989}, shown in Fig.~\ref{weinberg}:  According to standard physics, the vacuum has an enormous energy density $\rho _{vac}$. A typical  positive contribution is the zero-point energy of the electromagnetic field, and a typical negative contribution arises from Higgs condensation. All the various contributions are determined independently and there is no reason why they should cancel.

Again according to standard physics, $\rho _{vac}$ should act as a gravitational source -- effectively an enormous cosmological constant $\Lambda _{vac}$. It should then have an enormous effect on the curvature of spacetime, roughly 120 orders of magnitude larger than is compatible with observation (with the Planck scale providing a natural cutoff).

In the past few decades, Weinberg and others have adopted the point of view that, of all proposed solutions to this problem, the only acceptable one is the anthropic bound that he obtained in 1987~\cite{weinberg-1987}. However, anthropic reasoning -- see \ref{subsec:subsec17} -- is not universally accepted by the physics community. And it is likely that essentially all professional physicists would prefer a nonanthropic explanation of this largest of all discrepancies between standard theory and observation.

So despite decades of attempts by the best minds in theoretical physics, there is no truly convincing solution to this problem, and it may be signaling the need for a revolutionary conceptual breakthrough.

\subsection{\label{subsec:subsec1a}The dark energy problem}

\begin{figure}[htbp]
\centering
\includegraphics[bb=0 0 360 380, width=3.5in]{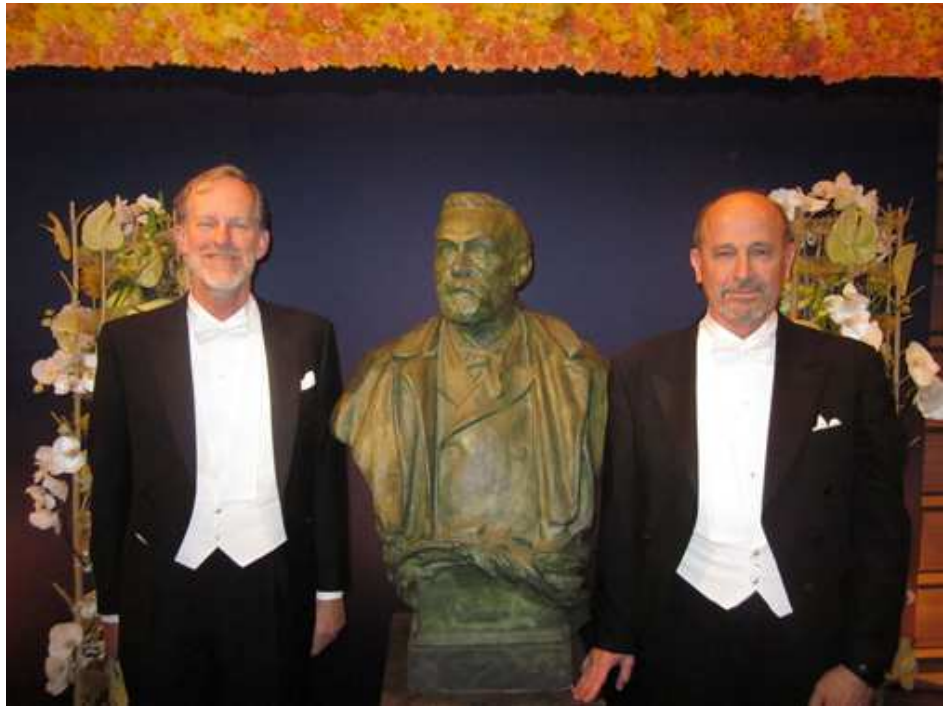}
 \caption{The acceleration in the expansion of the universe was discovered by the two groups of Refs. \cite{riess} and \cite{perlmutter}. The first of these groups was initially organized by Nicholas Suntzeff, on the right, together with Brian Schmidt. Mark Phillips, on the left, also played a leading role in this discovery. All members of the High-z Supernova Search Team and the Supernova Cosmology Project shared the 2007 Gruber Cosmology Prize. Here Nick Suntzeff and Mark Phillips are shown during the 2011 Nobel Week in Stockholm, where (in accordance with the traditional limitation to three recipients) the Physics Nobel Prize for the discovery went to Brian Schmidt, Adam Riess, and Saul Perlmutter. Credit: Nicholas Suntzeff.
\label{suntzeff}}
 \end{figure}
\begin{figure}[htbp]
\centering
\includegraphics[bb=0 0 360 220, width=5.5in]{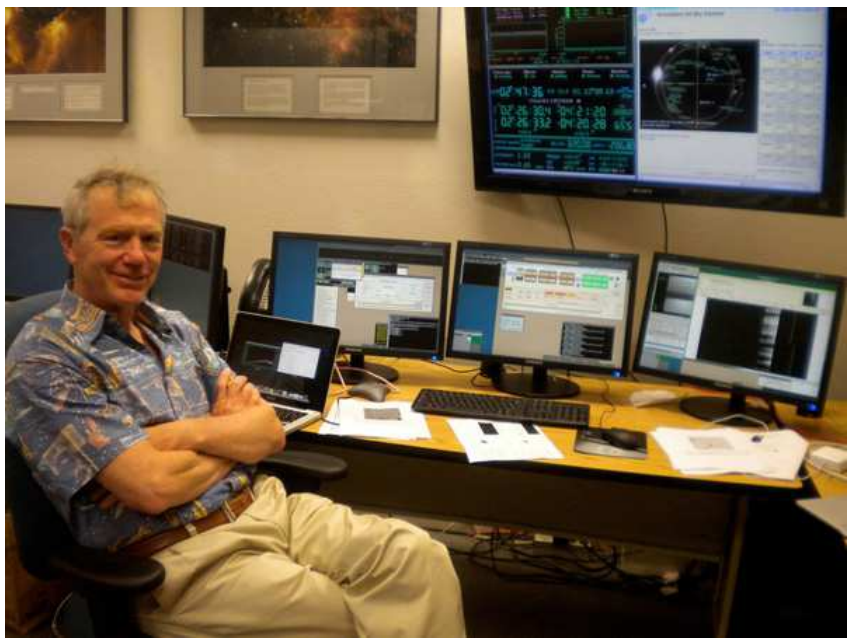}
 \caption{Robert Kirshner mentored a number of the High-z Supernova Search Team members, and two of the three Nobel recipients for the discovery of the accelerating universe, Brian Schmidt and Adam Riess, were his Ph.D. students. In this 2014 photo, he is observing supernovae at the Keck Observatory on Mauna Kea in Hawaii. Credit: Jayne Loader. \label{kirshner}}
 \end{figure}
\begin{figure}[htbp]
\centering
\includegraphics[bb=0 0 360 480, width=3.2in]{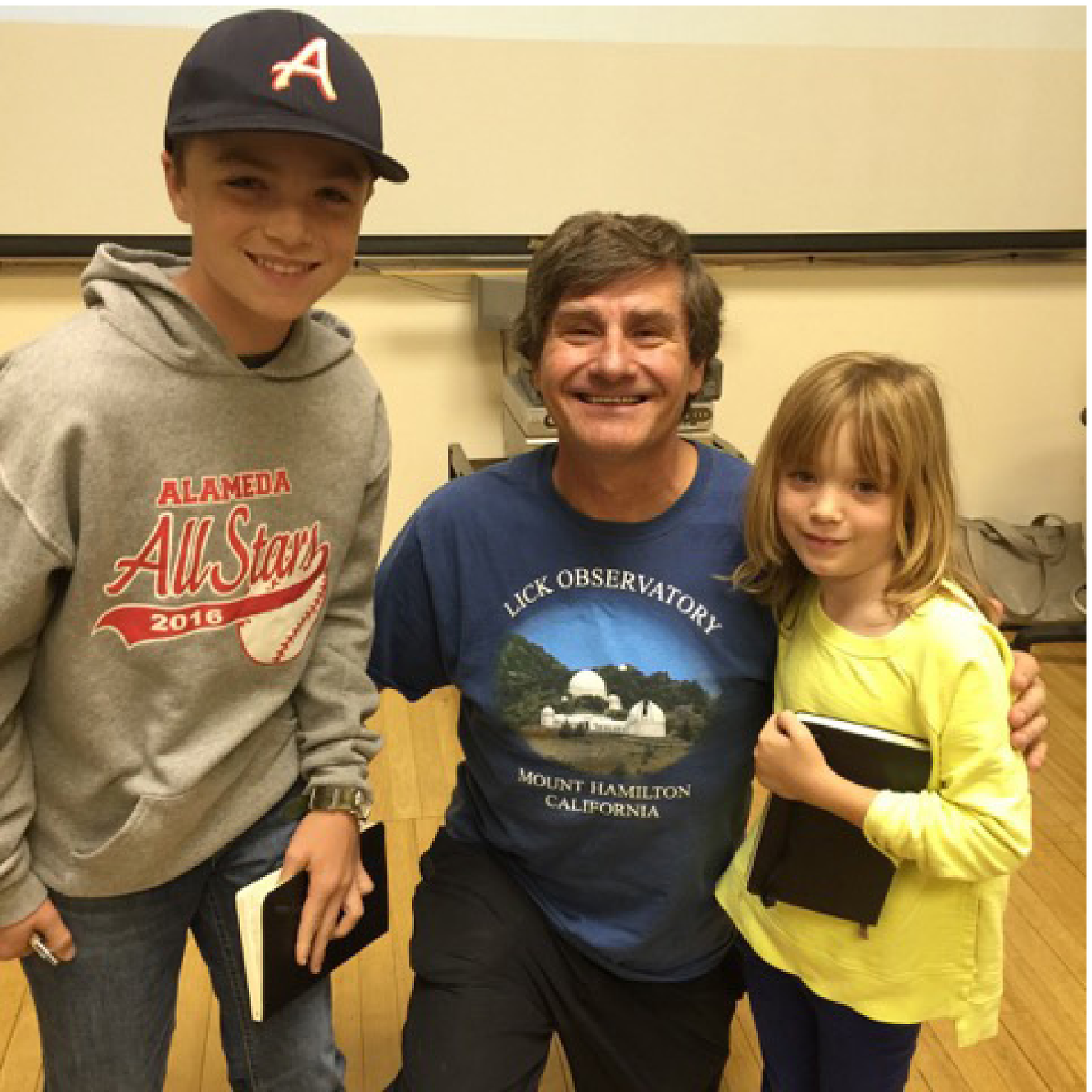}
 \caption{Alex Filippenko is the only person who was a member of both the High-z Supernova Search Team and the Supernova Cosmology Project. Here he is shown with Dahlia Smith, age 7, and  Scott Rapposelli, age 12, after his presentation on solar eclipses -- where they had taken careful notes. Credit: Erin Kelley- Smith.
\label{filippenko}}
 \end{figure}

In 1998 two groups, who had set out to measure the expected deceleration in the expansion of the universe (resulting from the gravitational attraction of ordinary matter),  instead made an astounding discovery: The universe has been \textit{accelerating} in its expansion for the last few billion years~\cite{riess,perlmutter}. Some of the members of one group are shown in Figs.~\ref{suntzeff}, \ref{kirshner}, and \ref{filippenko}. Increasingly accurate measurements have shown that the cause of this acceleration -- commonly called ``dark energy'' -- appears to have the same properties as a relatively tiny cosmological constant $\Lambda $~\cite{acc-review-2013}. 
An effectively repulsive gravitational force can arise in general relativity because both the pressure and the energy density act as sources of gravity. There are two simple alternatives if the origin of cosmic acceleration does have the same properties as a cosmological constant: Either it is a vacuum energy acting as a source term (and is thus truly a dark energy), or else it is instead a fundamental  cosmological constant (in which case ``dark energy'' is essentially a metaphor). However, the division between vacuum energy and bare cosmological constant will remain arbitrary until there is an accepted theory or nongravitational experiment that provides a definite value for the vacuum energy~\cite{fulling}. 
Again, this problem has been addressed in a vast number of papers, and the lack of a convincing explanation seems to indicate that our current understanding of gravity requires profound revision.

\subsection{\label{subsec:subsec2}Regularization of quantum gravity}

\begin{figure}[htbp]
\centering
\includegraphics[bb=0 0 360 200, width=6in]{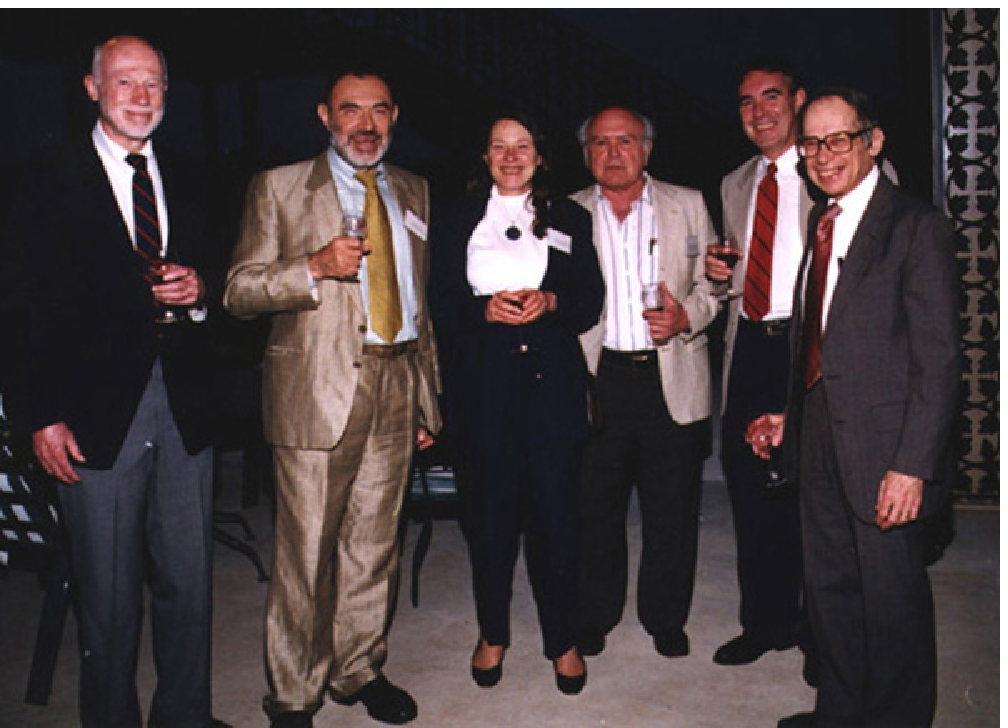}
 \caption{The theorists shown in this figure are among the major leaders in the long quest for a quantum theory of gravity (and fundamental understanding of other forces and particles). From left to right: Bryce DeWitt, who obtained the Wheeler-DeWitt equation for the ``wavefunction of the universe''; it is based on the ADM classical description of gravity developed by Arnowitt, Deser, and Misner, two of whom are also shown in this photo. Stanley Deser, co-discoverer of supergravity. Mary K. Gaillard, a leader in supersymmetry, superstrings, and related areas of theoretical physics. Bruno Zumino, co-discoverer of supersymmetry and supergravity. Michael Duff, a leading string theorist. Richard Arnowitt, co-inventor of 
the minimal supergravity model used in many analyses of experiment and observation.
\label{gravity}}
\end{figure}
It is a near consensus that gravity should ultimately be described by quantum mechanics, like the other fields of nature. This was already recognized by Einstein in 1916. But attempts to quantize gravity perturbatively were found to exhibit severe divergences as the cutoff in the calculations approaches the Planck energy of about $10^{16}$ TeV.

The history of the extremely large and varied number of efforts to quantize gravity is summarized by Rovelli~\cite{rovelli}, and some of the many important contributors in this enterprise are shown in Fig.~\ref{gravity}. The quantum Wheeler-DeWitt equation~\cite{dewitt} for the ``wavefunction of the universe''  follows from the classical 3+1 decomposition of general relativity by Arnowitt, Deser, and Misner~\cite{ADM} plus the work of many others, including Dirac.

But standard Einstein gravity has so far resisted all attempts at consistent quantization, because the coupling constant $\ell_P$ has the dimension of length (whereas the coupling constants of nongravitational forces are dimensionless), and this turns out to lead to extremely rapid divergences in a perturbative calculation of quantities like scattering cross-sections if there is not a fundamental cutoff in energy near $\ell_P^{-1}$.

There are two well-known attempts to quantize gravity by imposing such a cutoff: In string theory~\cite{witten,polchinski,beckers,schwarz,greene,kaku,randall}, the world-line of a particle is replaced by the world-sheet of a string, so that the intersections of lines in Feynman diagrams (which produce divergences) are broadened into the intersections of sheets (which may be e.g. cylindrical). There are many further extensions which are still under development, involving e.g. branes and a potentially unifying and all-encompassing M-theory. In loop quantum gravity~\cite{rovelli}, spacetime has a ``granularity''. Unfortunately, both of these (mathematically very appealing) approaches remain largely theories of gravity alone, in the sense that, after decades of brilliant work by many groups, they have still not managed either to make convincing contact with the rest of physics (including the Standard Model) or to make predictions that are experimentally testable.

So quantum gravity remains an extremely challenging and unsolved problem.

\subsection{\label{subsec:subsec2a}Black hole entropy and thermodynamics}

\begin{figure}[htbp]
\centering
 \includegraphics[bb=0 0 207 720, width=1.2in]{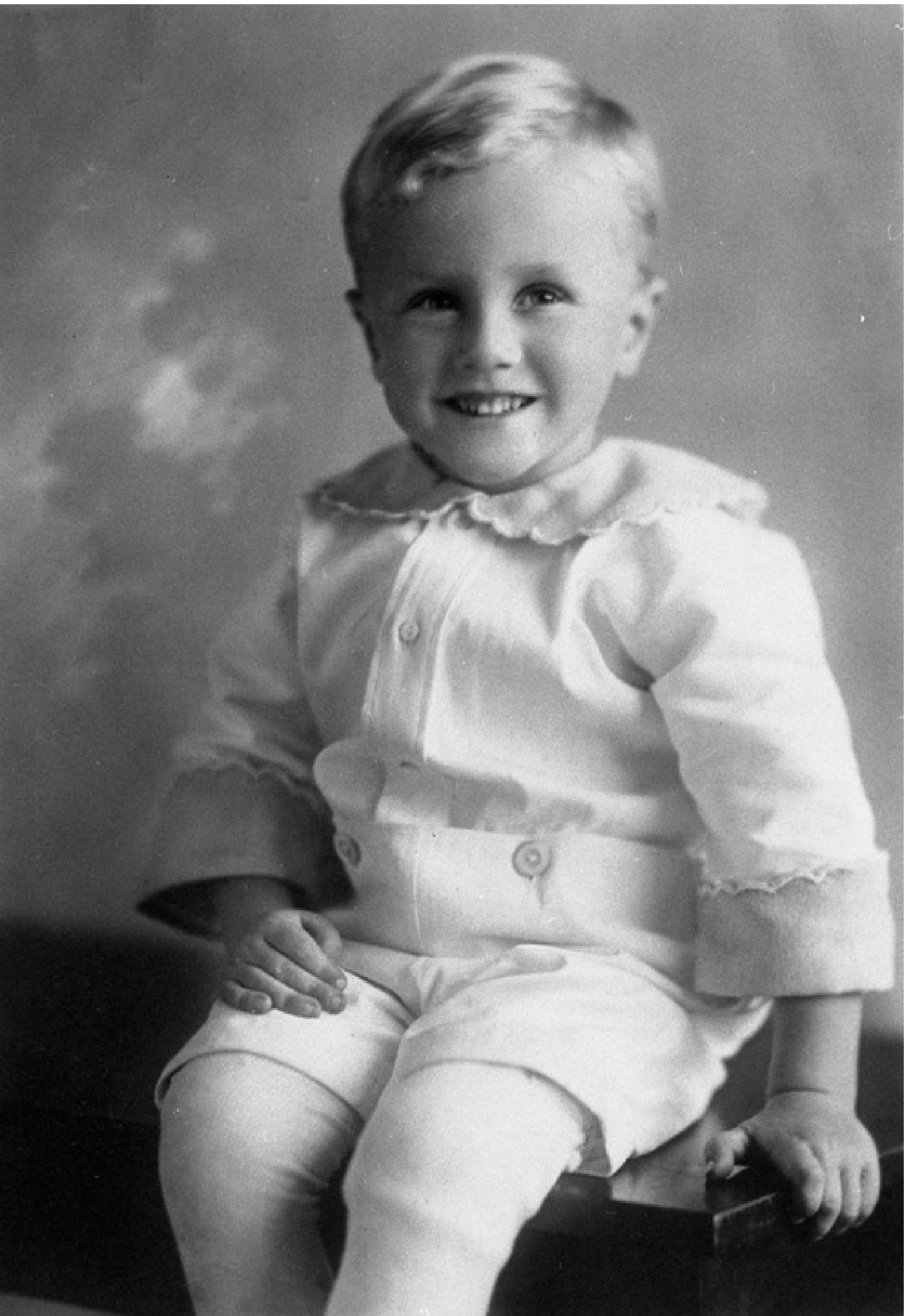}
 \caption{John Wheeler invented the terms \textit{black hole}, \textit{wormhole}, \textit{quantum foam}, and \textit{neutron moderator}, and made many important contributions to gravitational and nuclear physics. He is shown here at age 4. 
Credit: AIP Emilio Segre Visual Archives, Wheeler Collection.
\label{wheeler}}
 \end{figure}
Black holes were named by John Wheeler, shown in Fig.~\ref{wheeler}. They are now understood to play  central roles in astrophysics~\cite{thorne}, with both star-sized and supermassive black holes confirmed by a wealth of observational data. 

\begin{figure}[htbp]
\centering
 \includegraphics[bb=0 0 2648 1936, width=0.8\textwidth]{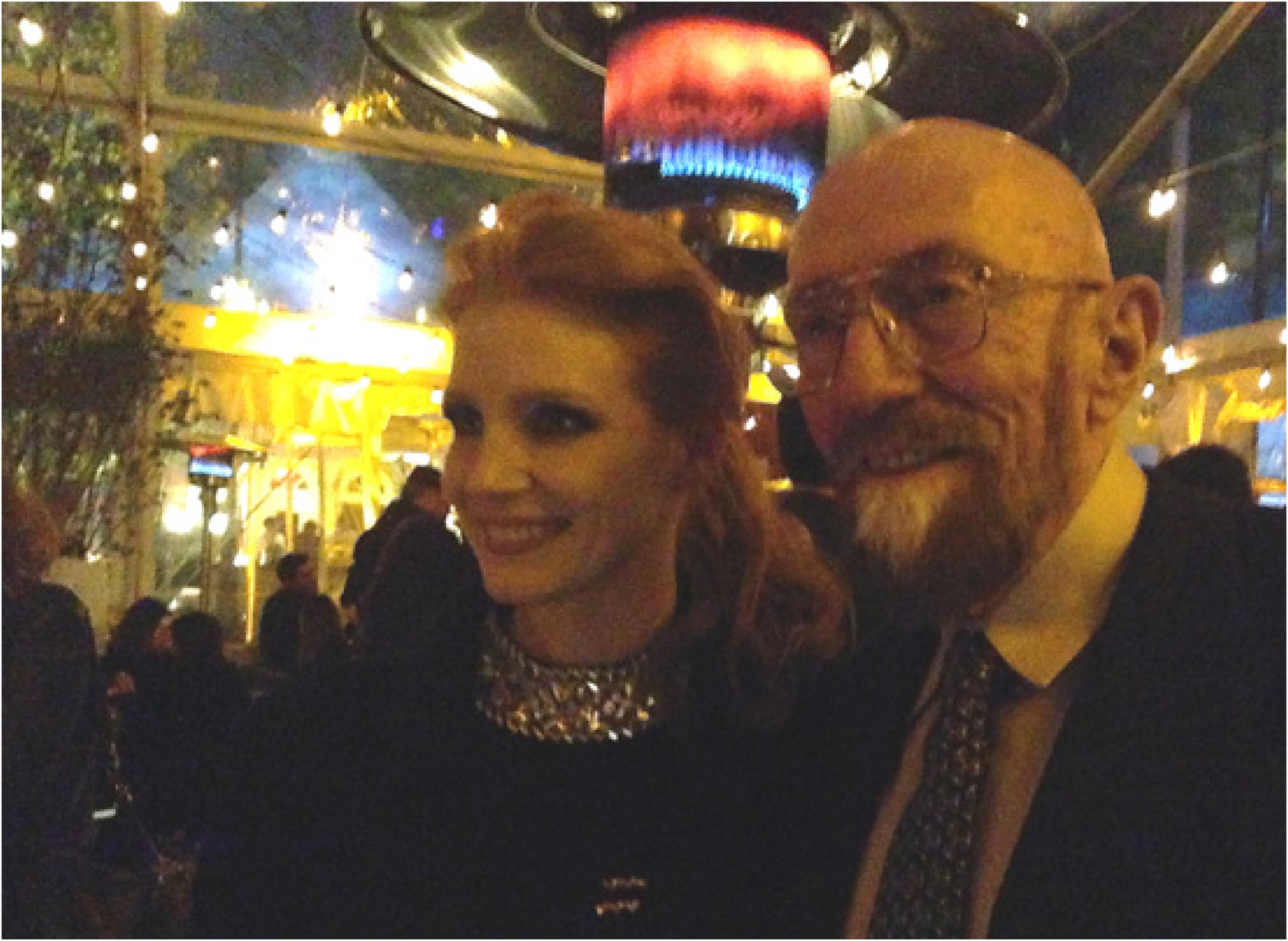}
 \caption{Kip Thorne with actress Jessica Chastain at the world premiere of \textit{Interstellar}. 
This movie was inspired by Kip Thorne and his producer friend Lynda Obst. He was also Executive Producer and Science Consultant, and wrote a book explaining the science in the film~\cite{Interstellar}. Earlier he had introduced wormholes into science fiction, via a novel of Carl Sagan -- \textit{Contact} -- which was made into a movie produced by Lynda Obst. 
Credit: Kip Thorne.
\label{Thorne1}}
 \end{figure}

\begin{figure}[htbp]
\centering
 \includegraphics[bb=0 0 360 310, width=4in]{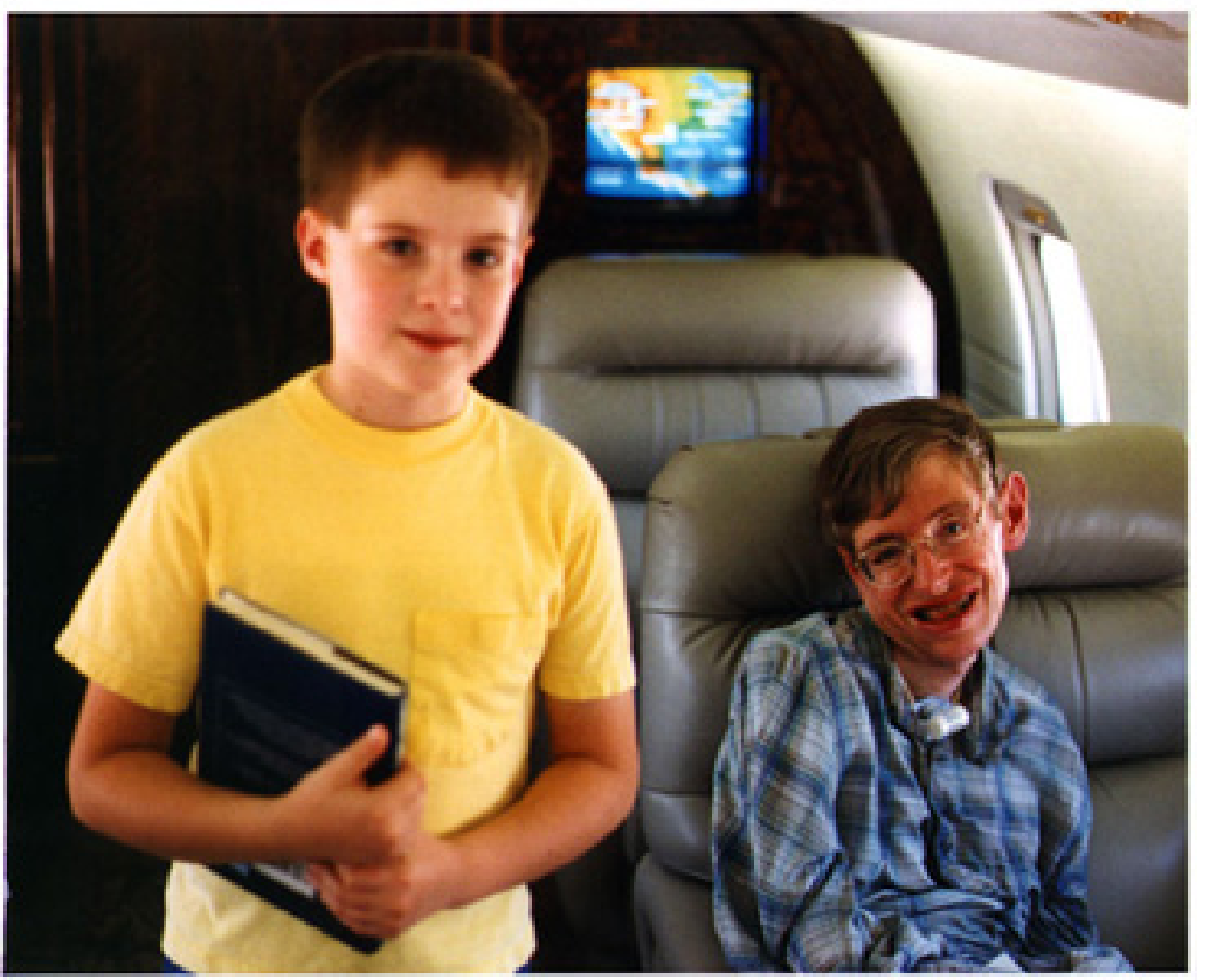}
 \caption{Stephen Hawking with a young fan in 1995. The book is \textit{A Brief History of Time}, autographed with the author's thumbprint. The plane is an EDS corporate jet based in Dallas, which had flown in from Cambridge, England. \label{Hawking}}
 \end{figure}

But controversies remain about the meaning of the black hole entropy and radiation derived by Jacob Bekenstein and Stephen Hawking, as is already clear from the series of famous bets~\cite{bets,susskind}  involving Hawking and Kip Thorne, shown in Figures~\ref{Thorne1}, \ref{Hawking}, \ref{Hawking-emu}, \ref{Thorne2}, and \ref{Thorne3}. The only clear winner so far is Thorne, although Hawking has conceded his part of a bet to  John Preskill for reasons that are not convincing to most experts.
The Bekenstein-Hawking entropy is given by 
\begin{eqnarray}
S_{BH}=\frac{1}{4}\frac{A}{\ell_{P}^2}
\label{2.4}
\end{eqnarray}
and the Hawking temperature (for quantum radiation of particles and antiparticles) by
\begin{eqnarray}
T_{H}=\frac{\kappa}{2 \pi}
\label{2.5}
\end{eqnarray}
where $A$ is the surface area and $\kappa$ the surface gravity of a black hole. Here $\ell_{P}$ is the Planck length of about $1.6 \times 10^{-25}$ \AA\ and the second equation is given in the units  often used by gravitational theorists: $\hbar=c=G=k=1$, where $\hbar= h/2 \pi$, $h$ is Planck's constant, $c$ is the speed of light, $G$ is the gravitational constant, and $k$ is the Boltzmann constant. 

It is clear that these quantities are closely related to both gravity and quantum mechanics, but a fundamental mystery is why the entropy should be proportional to area rather than volume, as is the case for other physical systems. 

\begin{figure}[htbp]
\centering
 \includegraphics[bb=0 0 360 540, width=4in]{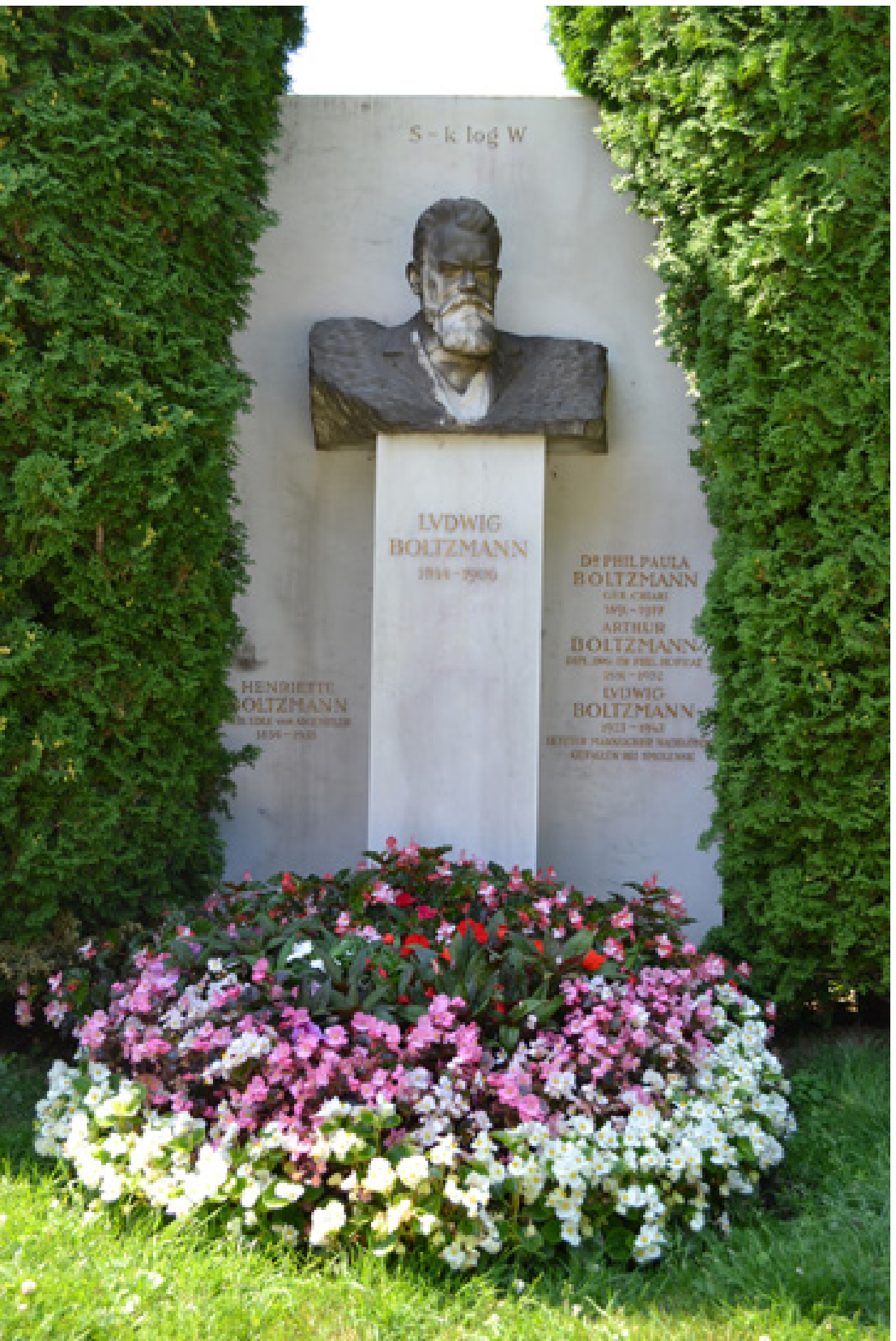}
 \caption{Tombstone of Ludwig Boltzmann in Vienna, showing his fundamental formula $S=$ $k$ log $W$, where $S$ is the entropy, $k$ is Boltzmann's constant, and $W$ is the number of available microstates. \label{boltzmann}}
 \end{figure}
One would like to start with the fundamental expression for entropy shown in Fig.~\ref{boltzmann},
\begin{eqnarray}
S_{BH} = k \log W_{BH} \; ,
\label{2.6}
\end{eqnarray}
\begin{figure}[htbp]
\centering
 \includegraphics[bb=0 0 360 300, width=4in]{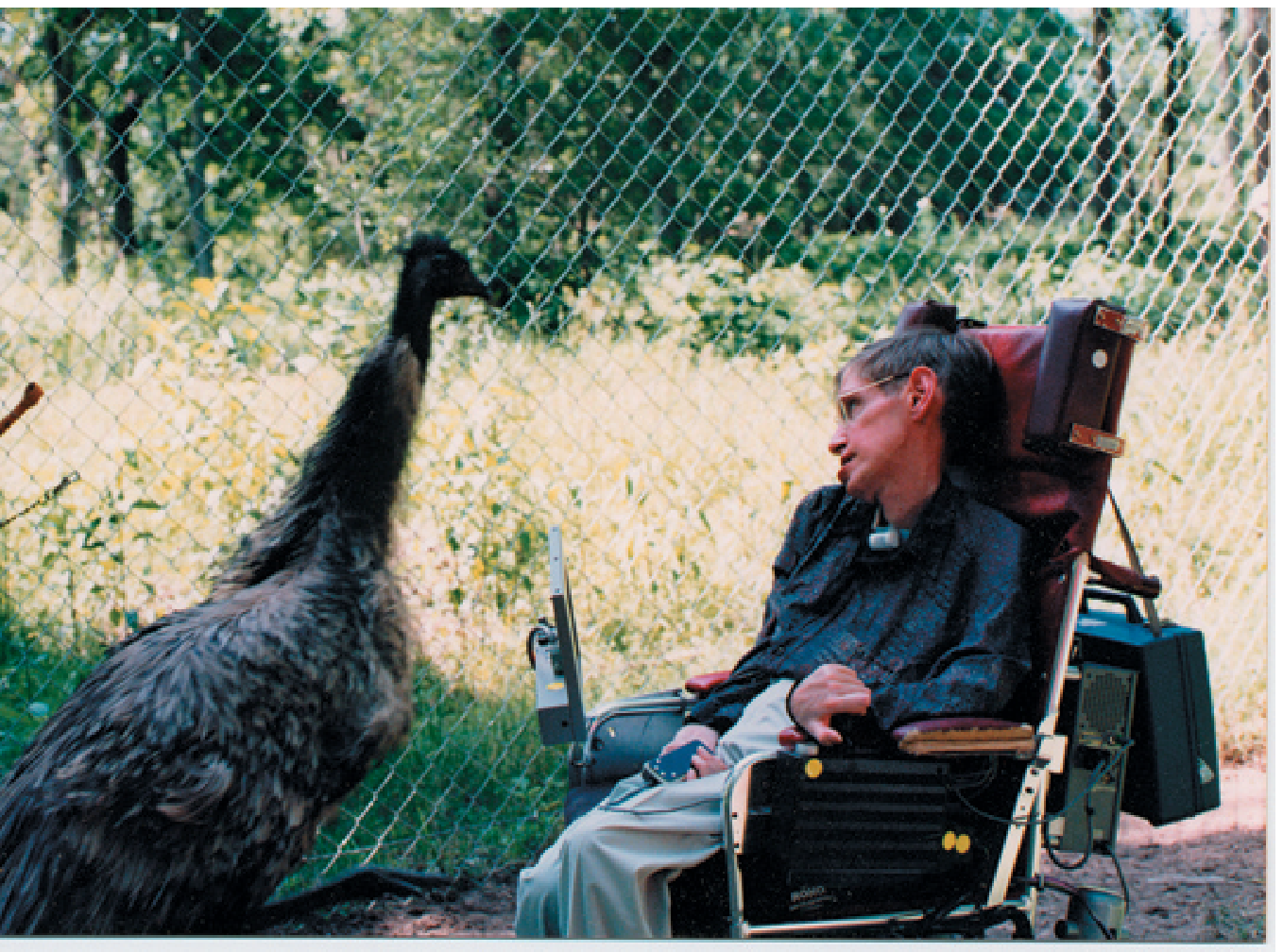}
 \caption{Stephen Hawking plays tag with an emu on a Texas ranch. Credit: Hans Schuessler. 
\label{Hawking-emu}}
 \end{figure}
(where ``BH'' represents ``black hole'' or `` Bekenstein-Hawking'')
and then obtain (\ref{2.4}) by counting the number of available microstates $W_{BH}$. There have been many attempts in this direction in string theory, in loop quantum gravity, and in various models, but so far it has not proved possible to derive (\ref{2.4}) for the general case of real black holes in 4-dimensional spacetime, or even for the simplest case of a static Schwarzschild black hole. The simplicity of (\ref{2.4}), and the complexity of the purported derivations starting with quite different theories, demonstrate that real understanding has not yet been achieved.

So a convincing interpretation of the black hole entropy (\ref{2.4}) remains a leading challenge for the theoretical community.

\subsection{\label{subsec:subsec2b}Black hole information processing} 

There are two possibilities for black hole thermodynamics:

If (as in ordinary thermodynamics) it is only a statistical description at the macroscopic level, essentially reflecting the ignorance of an observer, then there may be only an apparent loss of information when objects are incorporated into the black hole and later emerge as Hawking radiation. In this case there may be a deeper microscopic description (yet to be convincingly found) in which the time evolution is fully deterministic (and unitary) and no information is truly lost. E.g., if a mountain-size distribution of well-defined matter collapses to form a microscopic black hole, and later (after billions of years of the evaporation of mass through Hawking radiation) the black hole disappears in a final release of energy, there would then be subtle traces in the radiation which could (in principle) still be used to determine the detailed features of the original matter.

On the other hand, if the entropy and temperature are fundamental (rather than statistical) features of a black hole, determined directly by gravity and quantum mechanics alone, then the original matter (which collapsed to form the black hole) would have its detailed nature obliterated within the event horizon. In this case the original information would be lost, perhaps through quantum gravity processes to be understood in the future.

There are other ways of formulating the black hole information paradox, but it is clear that this issue is closely related to those of topics \ref{subsec:subsec2} and \ref{subsec:subsec2a} above, and that a definitive resolution appears to await a deeper understanding of quantum gravity. The various attempts to resolve this issue -- described by phrases like holographic principle, firewall, and black hole memory -- have been well covered in even the popular media, but none has so far proved convincing.

\subsection{\label{subsec:subsec3}Cosmic inflation (or an inflation-like scenario)}

Inflation is a postulated exponential expansion of the very early universe, perhaps by a factor of $10^{n}$ with $n \sim 40$, when its age was roughly $10^{-32}$ second. It was proposed to explain several features of our observable universe~\cite{guth,kolb,linde}, including its large-scale flatness and isotropy, and it predicts the nearly scale-free fluctuations in the cosmic microwave background radiation (CMB) that have been observed in increasing detail by the COBE, WMAP, and Planck missions~\cite{planck}. In this picture, extremely tiny quantum fluctuations were enormously stretched, yielding the variations in primordial radiation and galactic structure that are now spread across the sky.
\begin{figure}[htbp]
\centering
 \includegraphics[bb=0 0 1250 400, width=2.4\textwidth]{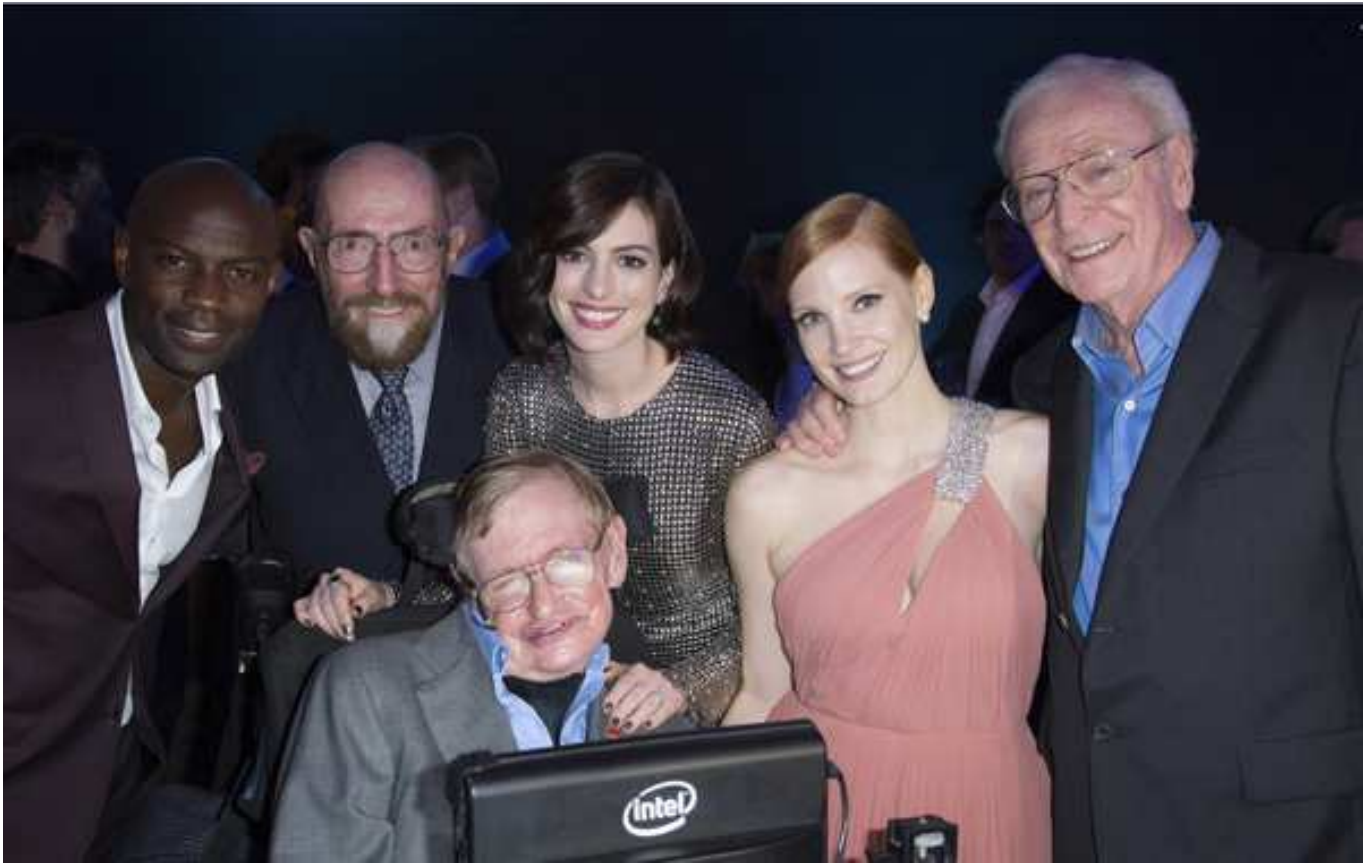}
 \caption{Kip Thorne and Stephen Hawking with actors of the movie \textit{Interstellar}: David Gyasi, Anne Hathaway, Jessica Chastain, and Michael Caine. 
Credit: Kip Thorne.
\label{Thorne2}}
 \end{figure}
A first major question, to be answered by observations, is whether there is direct evidence for inflation. A second, to be answered by both theory and observations, is the origin of inflation (if it is indeed validated). There are currently a vast number of competing models, many of which are highly speculative, and none of which is fully convincing. 

It is possible (and perhaps likely) that a definitive inflation-like scenario will be achieved only after a more general conceptual breakthrough.

\subsection{\label{subsec:subsec4}Cosmological survival of matter (and not antimatter)}

If the Standard Model were strictly obeyed, there should have been an essentially complete annihilation of matter and antimatter in the early universe, leaving only photons. So some extension of standard physics is needed to explain why instead there is a small amount of residual matter which makes up the familiar objects in ordinary experience and astronomy. The only alternative would be an extreme and unnatural fine-tuning in the initial state of the universe.

\begin{figure}[htbp]
\centering
 \includegraphics[bb=0 0 360 510, width=0.5\textwidth]{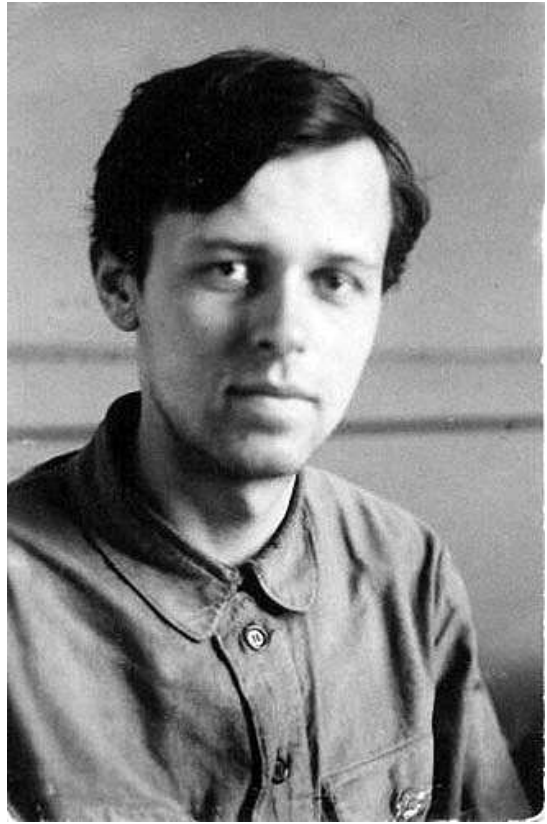}
 \caption{Andrei Sakharov, shown here in 1943, was the principal designer of the Soviet thermonuclear bomb in the years following World War II. But not long afterward he led the campaign against nuclear proliferation and atmospheric nuclear tests, and in favor of political reform. Labeled ``Domestic Enemy Number One'' by the head of the Soviet KGB, he was then subjected to sustained pressure, intimidation, threats, internal exile, and physical abuse, before he was released by Mikhail Gorbachev in 1986, a few years before the fall of the Soviet Union. He received the Nobel Peace Prize in 1975. Among his many contributions to applied and fundamental physics are the Sakharov criteria for baryogenesis, which are key to understanding how matter survived after the extreme conditions of the Big Bang.
Credit: Wikimedia Commons.
\label{Sakharov}}
 \end{figure}

This has been a major area of investigation -- exploring possible origins of baryogenesis or leptogenesis, in which physical processes beyond the Standard Model ultimately produce a sufficiently large asymmetry between matter and antimatter. In the case of baryogenesis, the required criteria were published in 1967 by Sakharov, shown in Fig.~\ref{Sakharov}: nonconservation of baryon number, C and CP violation, and interactions out of thermal equilibrium. Each has a simple explanation, once we define the terms: P is the parity operation, or roughly left $\longleftrightarrow$ right. C is the charge conjugation operation, or roughly particles $\longleftrightarrow$ antiparticles. T is time reversal, and CP or CPT means the product of the operations indicated. (i)~Since the baryon number is $+1$ for matter (e.g. a proton) and $-1$ for antimatter, and there is now more matter than antimatter, the baryon number must have somehow been changed by processes in the early universe if it was initially zero (as it would have been before any particles were produced). (ii)~Similarly, the symmetry between particles and antiparticles, described by C and CP, must be violated if the conversion of antimatter to matter is not to be counterbalanced by the conversion of  matter to antimatter. (iii)~Finally, there is the effect of CPT symmetry (see \ref{subsec:subsec11}), which would imply that there is still symmetry between matter and antimatter unless interactions occur out of thermal equilibrium, as in a rapidly expanding universe.

A key requirement of both baryogenesis and  leptogenesis is CP violation which is stronger than that already present in the Standard Model and which therefore requires new physics~\cite{trodden}.

 There are various theoretical proposals for the origin of the required CP violation, involving e.g. extra Higgs fields or supersymmetric models, but at the present none have received experimental support. So the dominance of matter over antimatter in the current universe (and even the fact that matter has survived) must still be counted as an outstanding fundamental problem.

\subsection{\label{subsec:subsec5}Composition of dark matter}

\begin{figure}[htbp]
\centering
 \includegraphics[bb=0 0 360 760, width=2in]{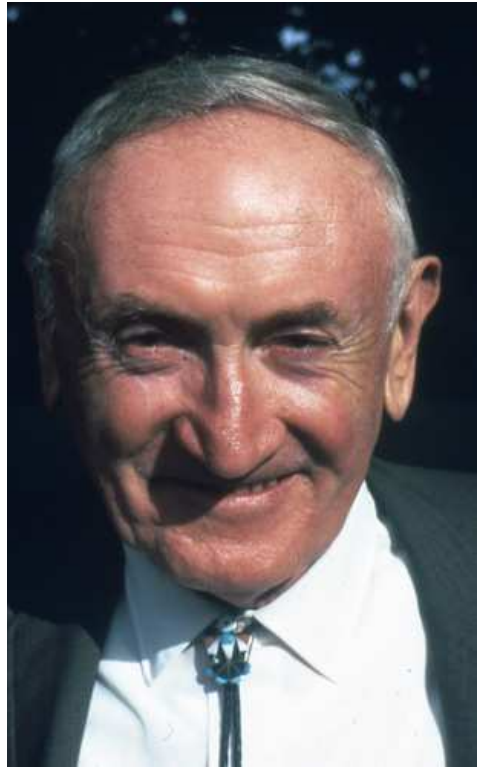}
 \caption{Fritz Zwicky at the International Astronomical Union meeting in Brighton, England, in 1970. During the 1930s he mapped out tens of thousands of galaxies, discovering that they tend to cluster, and obtaining evidence for what he named \textit{dark matter}. With his colleague Walter Baade he introduced the term ``supernova'' and proposed that supernovae signal the transition of ordinary stars into neutron stars, as well as being the origin of cosmic rays. He also predicted that galaxy clusters could act as gravitational lenses. These and others of his discoveries and ideas are at the heart of modern astronomy and cosmology.
Credit: AIP Emilio Segre Visual Archives, John Irwin Slide Collection.
\label{Zwicky}}
 \end{figure}

\begin{figure}[htbp]
\centering
 \includegraphics[bb=0 0 360 490, width=2.5in]{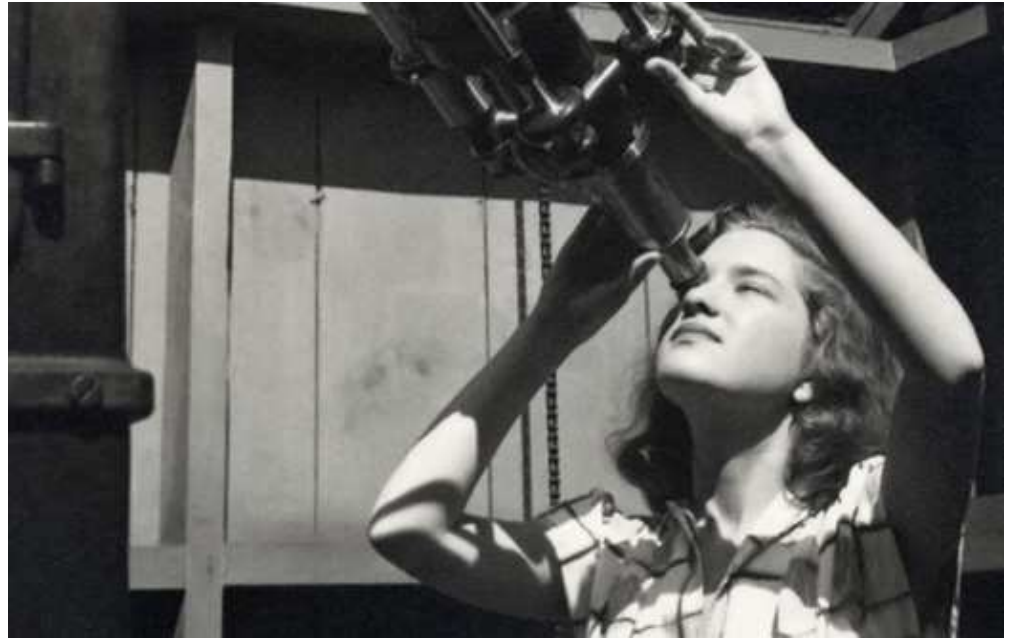}
 \caption{Vera Rubin, who provided a clear demonstration of the existence of dark matter by her measurement, with colleague Kent Ford, of the rotation speeds of stars around galaxies. A rapidly moving star requires a strong gravitational pull (from a large mass) to keep it in orbit.
Credit: Archives and Special Collections, Vassar College Library (unique identifier: Ph.f 6.93 Cooper Rubin, Vera).
\label{Rubin}}
 \end{figure}

The observations of Fritz Zwicky, in the early 1930s, and of Vera Rubin and her collaborators, beginning around  1970, indicated that the gravitational masses of galaxies are mostly due to something other than luminous matter. These two most central figures in the discovery of this invisible but gravitationally active  ``dark matter'' are shown in Figs.~\ref{Zwicky} and \ref{Rubin}.

The existence of dark matter has now been confirmed by many more recent observations, including those of the apparent separation of dark and luminous matter in colliding galactic clusters. The abundance of dark matter exceeds that of ordinary matter by a factor of five or six, and it has consequently played the dominant role in the formation of galaxies, galactic clusters, and larger-scale structures as the universe has evolved during the past 13.8 billion years.

This problem has been widely publicized, and is the current subject of many intense experimental and theoretical studies. For detailed discussion we defer to the readily available articles and papers, including  recent overviews~\cite{peter,ong,bergstrom,cline,freese}.
Despite increasingly powerful attempts to detect dark matter, through terrestrial collisions with atomic nuclei, emission of particles from extraterrestrial dark matter annihilations, and production of dark matter particles in accelerator laboratories, the composition of dark matter is still unknown. Among many candidates, the most popular are (i) the neutralinos (or other charge-neutral particles) of supersymmetric models (see \ref{subsec:subsec8a} below) and (ii) the axions predicted by the Peccei-Quinn theory for explaining why quark interactions do not exhibit strong CP violation (where this term is defined above in \ref{subsec:subsec4}). One of the most plausible arguments for weakly-interacting particles, like the lightest neutralino, is that the calculated cosmological abundance in the present universe is consistent with the observed abundance of dark matter.  
Characterizing the dark matter is certainly one of the best-known challenges confronting high-energy physics and astrophysics.

\section{\label{sec:sec3}Understanding and going beyond the Standard Model of particle physics}

The Standard Model has been remarkably successful in providing a highly quantitative description of all phenomena which do not involve gravity or neutrino masses~\cite{cheng,schwartz,pdg-2012}. But it remains in many ways a phenomenological model, with features that seem to call out for deeper explanation.

\subsection{\label{subsec:subsec6}Origin of family replication}

All of common experience is explained by the up quark, down quark, electron, and electron neutrino, but Nature has provided a second and even a third generation copy for each of these fermions. There is surely a profound reason for this repetition of family structure, but so far no convincing explanation has been given.

\subsection{\label{subsec:subsec9}Origin of particle masses}

Despite substantial efforts, no theory has proved capable of explaining the masses of fermions in the Standard Model, or equivalently their Yukawa couplings to the Higgs field. The top quark mass is especially mysterious, both because it is so large and because it appears to have a special value that is not far from the vacuum expectation value for a Standard Model Higgs field~\cite{chierici-sym,perez-sym}.

The discovery of neutrino masses~\cite{murayama,conrad,parke} has definitively taken physics beyond the Standard Model. In each generation of fermions, either an extra field must be added (for a  Dirac mass, like that of an electron or quark) or lepton number conservation must be violated (for a Majorana mass, which would mean that a neutrino is its own antiparticle). Explanations are needed for both the origin of neutrino masses and why they are so small. Each of these questions may find a satisfying answer in a grand unified theory (with e.g. an additional right-handed neutrino field and a seesaw mechanism to reduce observed neutrino masses), but there is still no universally accepted version of such a theory. 

At present the values of neutrino masses are unknown (since only differences in mass squared, $\Delta m^2$, are measured in neutrino oscillation experiments), and it is also not known whether the masses are Majorana or Dirac or both. In addition, just as for the other fermions, there is no convincing theory for the fundamental origin of these masses.

\subsection{\label{subsec:subsec8a}Supersymmetry and the hierarchy problems}

\begin{figure}[htbp]
\centering
\includegraphics[bb=0 0 360 240, width=5.5in]{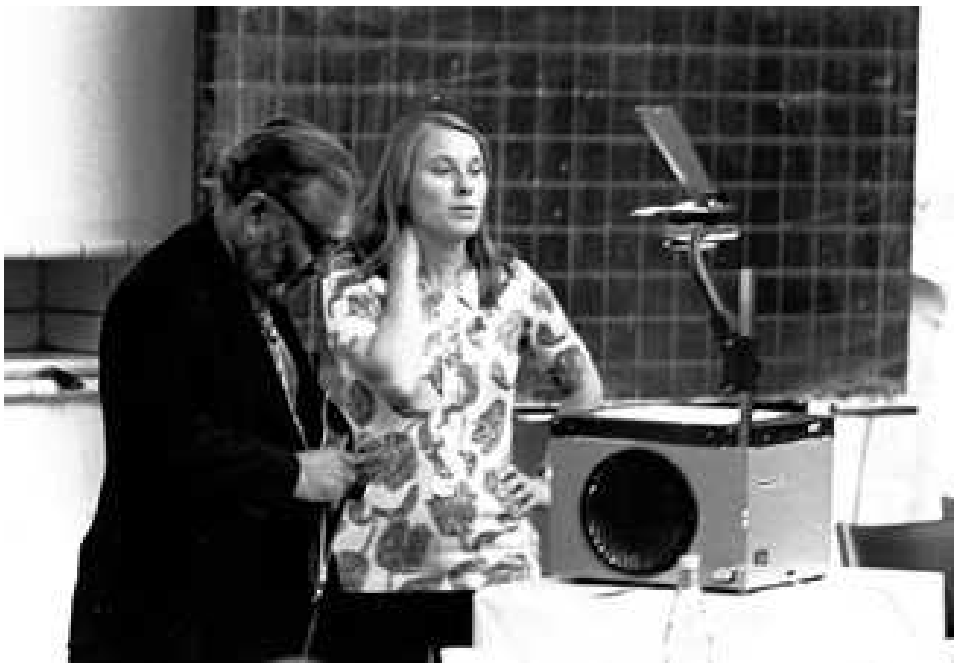}
 \caption{Mary K. Gaillard prepares to give a talk in a session chaired by Nobel Laureate Abdus Salam, at the 1976 neutrino conference in
Aachen. They have both been leading members of the large community of theorists exploring models of supersymmetry and unification of the forces of nature. Credit: Mary K. Gaillard.
\label{Gaillard}}
 \end{figure}

The recent observation of a Higgs boson~\cite{ATLAS,CMS,read,mariotti,ellis} may be regarded as completing the Standard Model, but it also appears to demand new physics, in order to protect the particle mass from quantum corrections that would increase it by perhaps 14 orders of magnitude or more~\cite{barbieri}. This is often called a hierarchy (or naturalness) problem. It is widely thought that the most plausible resolution of this problem is supersymmetry (susy)~\cite{martin,parker,nima}, which has been explored for the past four decades by many theorists, including those shown in Figs.~\ref{gravity} and \ref{Gaillard}. In a supersymmetric theory, for every standard matter particle (like electrons and quarks) there is at least one new force-carrying particle, and for every standard force-carrying particle (like photons, Z bosons, and Higgs particles) there is at least one new matter particle. In many supersymmetric theories, the lightest of the new matter particles is a stable particle with no electric charge -- the neutralino. This is one version of a weakly interacting massive particle (WIMP), which turns out to emerge from the early universe in about the right abundance to explain the dark matter observed in astronomy.
%

Other evidence for susy is the convergence of coupling constants mentioned in topic \ref {subsec:subsec10} and the fact that susy provides an extremely natural dark matter candidate, mentioned in topic \ref{subsec:subsec5}. However, the simplest supersymmetric models have been disconfirmed, and no convincing mechanism has yet been found either to break susy or to determine the many susy parameters. So there are several extremely compelling issues yet to be resolved by experiment and theory: validation or disconfirmation of susy; if there is in fact experimental validation, creation of a well-defined and realistic susy theory; and, within such a theory,  determination of the specific mechanism for susy breaking.

A second hierarchy (or naturalness) problem is why the electroweak scale is so tiny to begin with. The natural scales associated with the fundamental forces of nature -- the GUT and Planck scales -- are at about $10^{13} - 10^{16}$ TeV, so why are the masses of Standard Model particles many orders of magnitude smaller?

\subsection{\label{subsec:subsec10}Explanation of the fundamental grand unified gauge group}

Grand unification of the three nongravitational forces~\cite{kounnas,mohapatra,barger} is supported by (i)~the discovery of neutrino masses, since both Majorana and Dirac masses are natural in a grand unified theory (GUT), and (ii)~the realization that the fundamental charges for the electromagnetic, weak nuclear, and strong nuclear forces all converge to a common value at high energy when susy is included. However, the fundamental GUT gauge group has not yet been established. 

Furthermore, even if grand unification is eventually validated -- e.g. through observations of predictions like proton decay -- and the specific GUT theory is definitively determined, there is still a fundamental question to be answered: why this particular theory is in fact Nature's choice.

\subsection{\label{subsec:subsec11}Potential violation of Lorentz or CPT invariance}

\begin{figure}[htbp]
\centering
 \includegraphics[bb=0 0 360 470, width=3in]{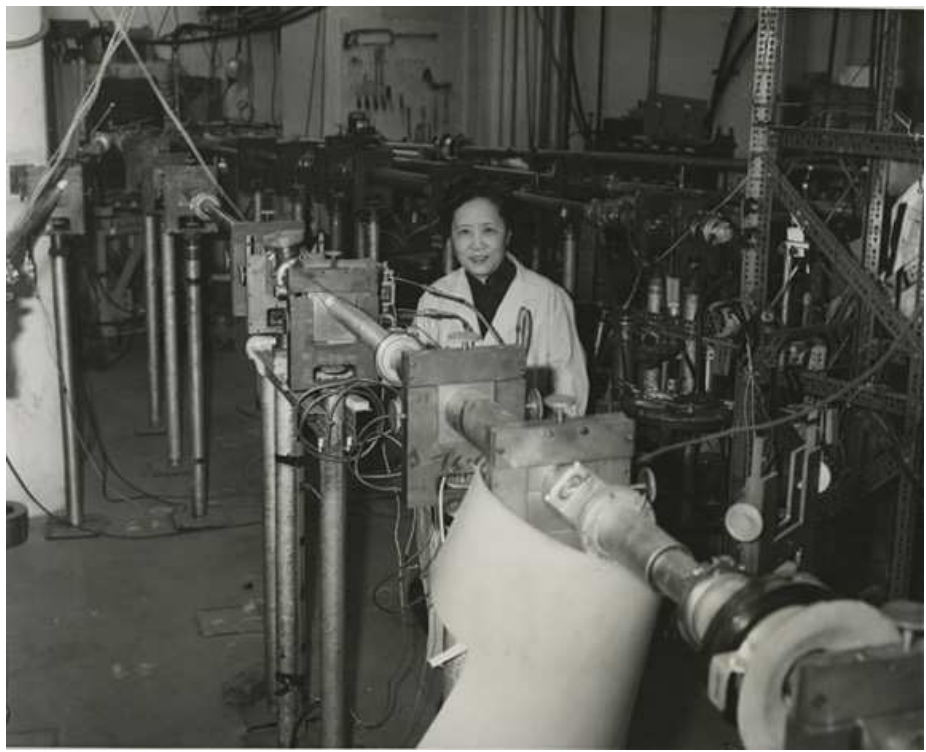}
 \caption{Experimental physicist Chien-Shiung Wu. She led the 1956 experiment that demonstrated parity violation -- violation of left-right symmetry -- by the weak nuclear interaction. This result, which confirmed a prediction of Chen-Ning Yang and Tsung-Dao Lee, came as a shock to many top scientists. But since then even CP, the combined operation of particle $\leftrightarrow$ antiparticle and left $\leftrightarrow$ right, has been found to be violated. This is explained by the existence of a third copy of the first generation of fermions in the Standard Model.
Credit: SIA Acc. 90-105 [SIA2010-1507] - Science Service, Records, 1920s-1970s, Smithsonian Institution Archives Image SIA2010-1507.
\label{Wu}}
 \end{figure}
\begin{figure}[htbp]
\centering
 \includegraphics[bb=0 0 380 280, width=6in]{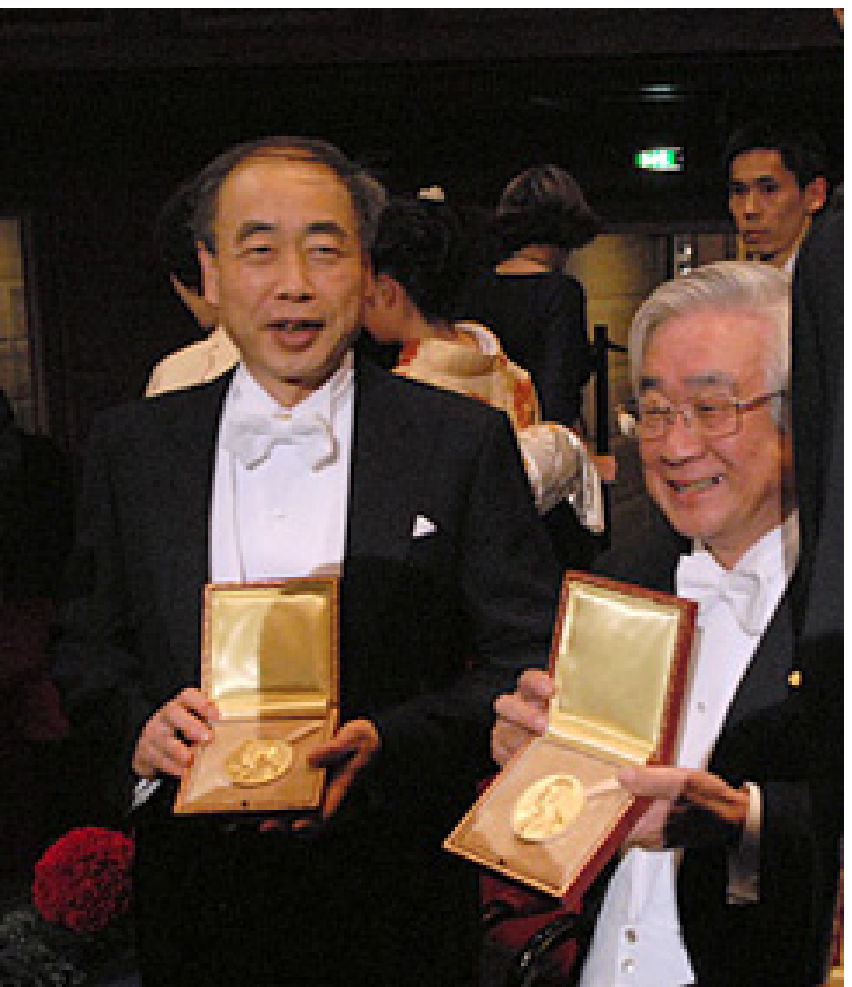}
 \caption{Makoto Kobayashi and Toshihide Maskawa display their Nobel Prize Medals after the awards ceremony in 2008. They showed that a third generation of matter particles implies the CP violation which had been discovered in 1964 by James Cronin and Val Fitch. A third generation was thus predicted, and its members were subsequently found, one by one, culminating in the discoveries of the top quark and tau neutrino at Fermilab. Credit: \copyright The Nobel Foundation, Photo: Hans Mehlin
\label{KM}}
 \end{figure}
The Standard Model violates P and CP symmetry at the most fundamental level, and also violates conservation of weak isospin and weak hypercharge (for particle interactions) after condensation of the Higgs field. The first observation of these symmetry breakings was demonstrated in an experiment led by Chien-Shiung Wu, shown in Fig.~\ref{Wu}.  

It is then natural to ask whether still more symmetries are broken either at a fundamental level (in a future theory which transcends the Standard Model) or because of further symmetry breakings (e.g. due to condensation of a vector or tensor rather than scalar field) or because of quantum fluctuations (such as ``spacetime foam'' at the Planck scale). So far no deviations from either Lorentz or CPT invariance have been observed~\cite{kostelecky,abdo}, but this is still an active area of investigation and violations of these most  fundamental symmetries may be quite subtle.

\subsection{\label{subsec:subsec14}Apparent marginality of  the Higgs self-coupling, and stability  of our universe}

It is remarkable that the recently discovered Higgs boson has a mass whose value implies that the fundamental self-coupling parameter in the Higgs potential 
\begin{equation}
V=-\mu ^2 h ^2+\lambda h ^4
\label{eq2.9}
\end{equation}
is very nearly equal to zero~\cite{perez-sym,altarelli,alekhin,degrassi} (if a Standard Model calculation is valid in this respect). According to this result, the Higgs condensate, and the universe as we know it, are only marginally stable. In fact, further calculations imply that our universe may be in a metastable phase, which eventually would undergo a transition to a more stable phase with very different properties‍.

This is a potentially very deep issue: what explains
\begin{equation}
\lambda \approx 0
\label{eq2.10}
\end{equation}
and is our universe in a stable phase or not?

\subsection{\label{subsec:subsec25}Quark confinement and related issues} 

The principle that quarks are always confined, and that all free particles have zero net color charge, is universally accepted, and is increasingly indicated by lattice QCD calculations, but it has never been rigorously proved. (The transition to a quark-gluon plasma at extremely high temperature is, of course, a separate matter.) 
\begin{figure}[htbp]
\centering
 \includegraphics[bb=0 0 360 600, width=2.3in]{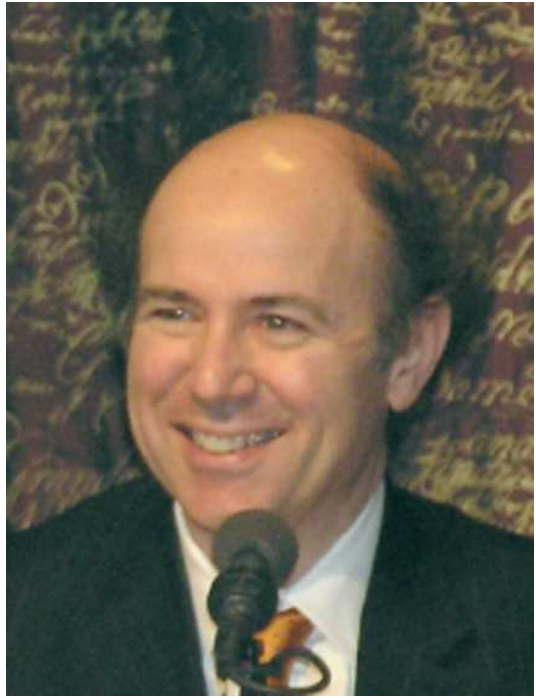}
 \caption{Frank Wilczek was awarded the Nobel Prize, with David Gross and David Politzer, for their demonstration of the ``asymptotic freedom'' of quarks at high energy or small distance, which makes it possible to perform perturbative calculations for the strong nuclear force at high energy. It is believed that the opposite behavior holds at low energy or large distance -- with quarks confined and unable to escape from e.g. protons or neutrons -- but this has not yet been rigorously proved (although it is increasingly indicated  by lattice gauge calculations). 
Credit: Betsy Devine.
\label{wilczek}}
 \end{figure}
This and related mathematical issues involving the strong nuclear force -- QCD -- are so deep and important that one is the subject of an unsolved Millennium Prize Problem~\cite{quigg-clay,millennium}.

Figure~\ref{wilczek} shows Frank Wilczek, a pioneer of QCD whose work has also covered many other deep issues in high energy and condensed matter physics. 

\subsection{\label{subsec:subsec25a}Phases of quantum chromodynamics and general systems with nonabelian gauge interactions} 

Various aspects of QCD are not well understood because of their inherent difficulties~\cite{deroeck,sjostrand}.

Since QCD processes are of central importance in collider experiments employing protons or ions, more precise treatments are needed of aspects like the formation of jets and parton distribution functions (or the ``internal landscape of the nucleons'').

The quark-gluon plasma is important in cosmology, and it appears to have been observed at the Relativistic Heavy Ion Collider (RHIC) and the Large Hadron Collider (LHC), where it is the main subject for the ALICE experiment. Its properties, which are being investigated in detail, have revealed surprises and new  insights and many new questions have arisen~\cite{schukraft,muller,QCD-transition,Saskia}. Regarding some of the unexpected features or mysteries, we quote Ref.~\cite{Saskia}: ``It is now widely believed that the creation of a deconfined phase of quarks and gluons has been achieved in high-energy heavy-ion collisions. However, the expectation of a weakly interacting (perturbative) system (as first ‘dreamed’ about in the 1970s) is contradicted by the data. The experimental observables have led us to conclude that the medium is strongly interacting, with a large degree of collectivity, even with small viscosity. Comparisons of hydrodynamic models to the data imply that the medium produced has a close to the conjectured minimum in viscosity to entropy density ratio.''

An additional intriguing topic is the complete phase diagram for QCD, and its applications in nuclear physics and astrophysics -- for example, in describing the interior structure of neutron stars. See, e.g., Fig. 1 of Ref.~\cite{Saskia} (taken from the U.S. Nuclear Science Advisory Committee 2007 Long Range Plan), which schematically depicts the basic features of the expected phase diagram.

And there is the related issue of the QCD vacuum, which is presumably filled with quark-antiquark condensates~\cite{cheng,schwartz}, even though this has not yet been proven. 

A still broader question may be of equal long-term interest: What are the emergent properties of multiparticle systems whose fundamental 2-body interaction is mediated by nonabelian gauge interactions?~\cite{Heinz} A specific example is the emergence of atomic nuclei, a profound problem that may come into reach of improved theoretical and computational techniques within even the next decade.

\subsection{\label{subsec:subsec12b}Additional undiscovered particles}

In the past, increasingly powerful accelerators or detectors have led to surprising discoveries of new particles, and this may happen again. Some have been postulated to solve problems -- for example, the axion, to explain why quantum chromodynamics (QCD) does not strongly violate CP invariance, and sterile neutrinos, to explain their possible observation in some neutrino oscillation experiments. Others are discussed primarily because they are  theoretical possibilities -- for example, additional fermions or bosons like those of the Standard Model.
\begin{figure}[htbp]
\centering
 \includegraphics[bb=0 0 360 540, width=2in]{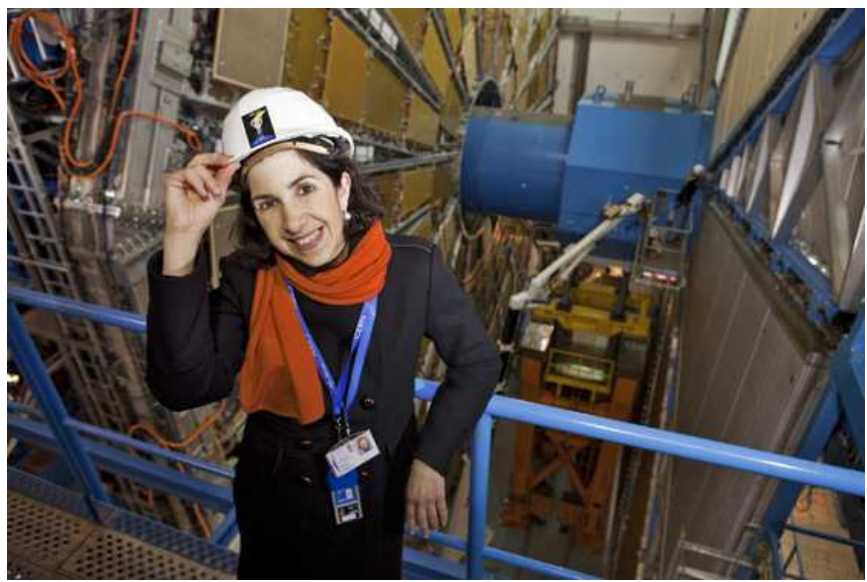}
 \caption{Fabiola Gianotti, current Director-General of CERN (European Organization for Nuclear Research).  What new discoveries lie ahead for the Large Hadron Collider, the world's most powerful scientific instrument? Credit: CERN.
\label{Gianotti}}
 \end{figure}
But the possibility of complete surprises in new experiments is always taken seriously, because our understanding of Nature is still quite incomplete. The open-minded search for new particles continues at  major accelerator laboratories such as CERN, whose Director-General, Fabiola Gianotti, is shown in Fig.~\ref{Gianotti}.

\subsection{\label{subsec:subsec12j}The unlimited future of astrophysics}

The cosmos is now known to be inhabited by many kinds of exotic objects. Whereas normal stars are supported against gravitational collapse by radiation pressure, white dwarfs are supported by the ``degeneracy pressure'' of electrons, which effectively repel each other because of the Pauli exclusion principle in space. Subrahmanyan Chandrasekhar, shown in Fig.~\ref{Chandra}, showed that the degeneracy pressure succumbs to gravity beyond about 1.4 solar masses. Neutron stars are similarly supported by the degeneracy pressure of neutrons and their constituent quarks. Those which are rapidly spinning can be observed as pulsars, discovered in 1967 by Jocelyn Bell Burnell, who can be seen in Fig.~\ref{Bell}. Stellar-size black holes can be observed by the emission of X-rays from their accretion discs. Supermassive black holes lie near the centers of almost all known massive galaxies. The cosmos is also filled with various kinds of particles and radiation from many sources. 

Given the fact that observational astronomy has yielded so many surprises in only the past few decades, there are surely many more discoveries waiting to be made. Among the vast number of possibilities that have been suggested by theorists, there are two intriguing possibilities for new kinds of stars that have not yet been observed: Population 3 stars are thought to have formed in the early universe and to have been composed almost entirely of hydrogen and helium. This would permit them to have much greater masses than the Population 1 (younger) and Population 2 (older) stars visible today. A very intriguing possibility is the ``dark stars'' proposed by Katherine Freese, shown in Fig.~\ref{Freese}, and her collaborators. The energy source for these objects would be dark matter annihilation rather than fusion reactions.

\section{\label{sec:sec3}The exotic behavior of condensed matter and quantum systems}

\begin{figure}[htbp]
\centering
 \includegraphics[bb=0 0 360 470, width=2.4in]{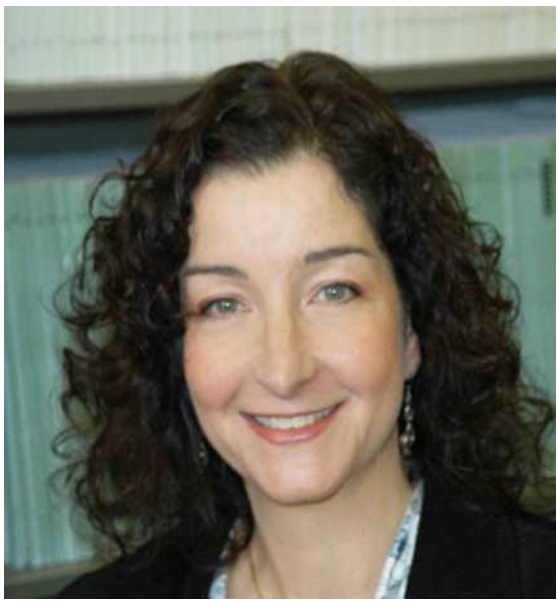}
 \caption{Laura Greene is Francis Eppes Professor of Physics and chief scientist at the National High Magnetic Field Laboratory. She is currently president-elect of the American Physical Society. Her research focuses on materials with strongly correlated electrons, including unconventional or novel superconductors. Credit: Laura Greene.
\label{Greene}}
 \end{figure}

\begin{figure}[htbp]
\centering
 \includegraphics[bb=0 0 360 740, width=2in]{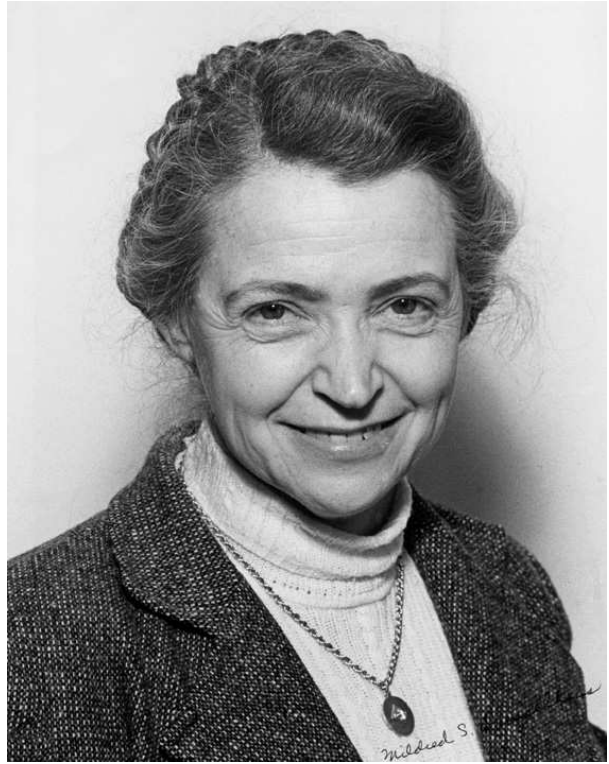}
 \caption{Mildred Dresselhaus has been called the ``queen of carbon science'', a vast field that spans several varieties of advanced materials with exotic properties (graphite, graphite intercalation compounds, fullerene molecules, carbon nanotubes, diamond, graphene, ...). She has received many honors and held many influential positions, including Institute Professor at MIT, director of the Office of Science at the U.S. Department of Energy, president of the American Physical Society, and president of the American Association for the Advancement of Science.
Credit: Photo by Calvin Campbell, Massachusetts Institute of Technology, courtesy AIP Emilio Segre Visual Archives, Gallery Of Member Society Presidents.
\label{dresselhaus}}
 \end{figure}

\begin{figure}[htbp]
\centering
 \includegraphics[bb=0 0 360 480, width=2.8in]{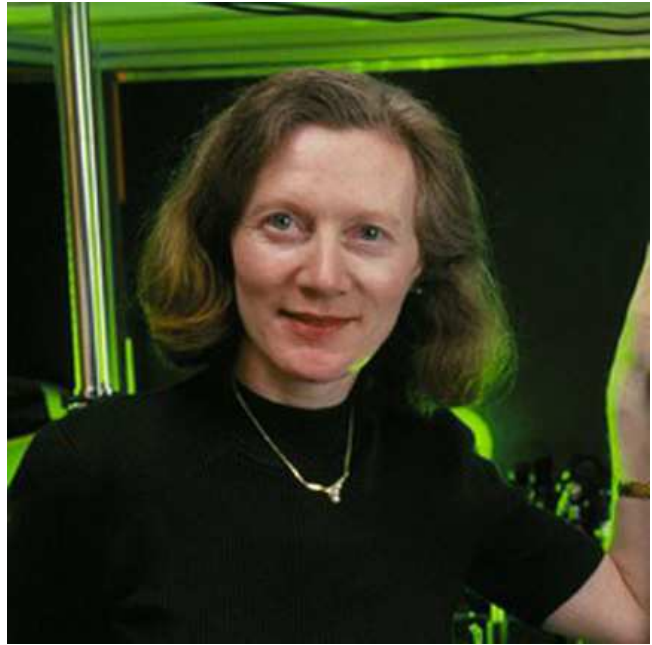}
 \caption{Margaret Murnane has pioneered many techniques in optical and x-ray science using new ultrafast laser and coherent x-ray sources, with potential applications in physics, chemistry, materials science, and engineering.  
Credit: University of Colorado.
\label{Murnane}}
 \end{figure}

\begin{figure}[htbp]
\centering
 \includegraphics[bb=0 0 360 670, width=2.5in]{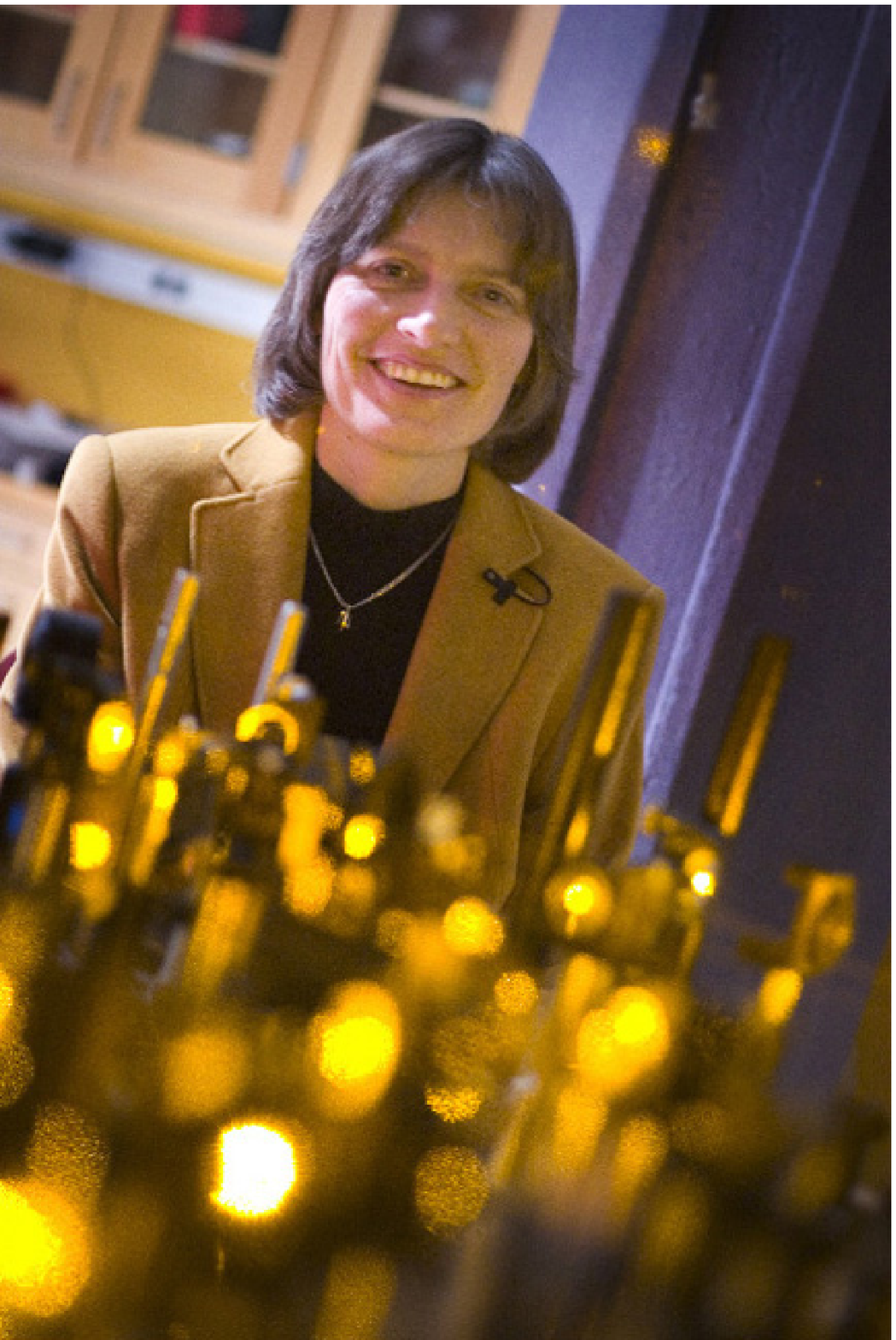}
 \caption{Lene Hau at a laser light optical table in her laboratory. She led a team which, by use of a Bose-Einstein condensate, succeeded in slowing a light beam to a few meters per second, and in 2001 was able to stop a beam completely. Later work led to the transfer of light to matter, then from matter back to light. This is a novel form of quantum control with potential applications in  quantum information processing and quantum cryptography. Credit: Justin Ide/Harvard News Office.
\label{Hau}}
 \end{figure}

Figures~\ref{Greene} - \ref{Hau} show some of the leading current researchers in these broad areas, who are working on problems like those considered below.

\subsection{\label{subsec:subsec12bb}What new forms of superconductivity and superfluidity remain to be discovered?}

At low temperature, bosons like $^4$He atoms can undergo Bose-Einstein condensation into a superfluid. Similarly, fermions can form pairs which also condense into a superfluid, or a superconductor if the fermions are charged~\cite{BCS}. There is currently a remarkable richness of known superfluids -- ranging from the superfluid phases of $^3$He to atomic Bose-Einstein condensates to the neutrons in a neutron star -- and superconductors, ranging from elemental metals to organics to ruthenates to heavy-fermion compounds to doped fullerenes to high-temperature superconductors -- the cuprates and newer iron-based materials.

The mechanism of superconductivity -- along with many other features -- is yet to be elucidated for high-temperature superconductors. And, given the many surprises in this area during even the last few years, it is likely that more major discoveries lie ahead.

\subsection{\label{subsec:subsec12dd}What new topological phases remain to be discovered?}

Topological insulators are another surprising discovery of recent years~\cite{kane}, following the earlier discoveries of the Kosterlitz-Thouless transition and the integer and fractional quantum Hall effects. There are various theoretical proposals for other topologically nontrivial phases and objects in condensed matter systems.

\subsection{\label{subsec:subsec12cc}What further properties remain to be discovered in highly correlated electronic materials?}

It is rather miraculous that a one-electron (or quasiparticle) picture works so well for many condensed matter systems. But correlation effects can lead to qualitatively new phenomena, and those mentioned above certainly do not exhaust all the possibilities.

\subsection{\label{subsec:subsec12ee}What other new phases and forms of matter remain to be discovered?}

The emergent properties of ordinary matter have been found to exhibit an amazing richness~\cite{Kohn}, with many other exotic phases discovered during the 20th and early 21st Centuries: various forms of magnetism, spatial structures (such as crystals and quasicrystals, charge density waves, spin density waves, pair density waves, and stripes in cuprates), 1- and 2-dimensional materials, nanostructures, soft matter (such as liquid crystals and polymers), and granular systems. 

Quantum phase transitions -- when parameters such as chemical doping, magnetic fields, or pressure are varied at zero temperature -- are currently an area of intense exploration. Quantum liquids, including the electron liquids in ordinary materials, are still not well understood, and the mere existence of any liquid phase is a nontrivial emergent property of matter~\cite{zangwill}. 

Turbulence in fluids is still regarded as a major unsolved problem, and understanding the Navier-Stokes equations is the subject of another Millennium Prize Problem~\cite{millennium}. It is also likely that there are more surprises lurking in more general nonlinear systems, involving, for example, chaos and nonequilibrium phase transitions.

Plasma has been described as the fourth state of matter, important in many areas of astrophysics and terrestrial applications. An old dream, still unfulfilled, is that a breakthrough in either magnetic confinement or inertial confinement will allow controlled fusion to become an almost unlimited source of usable (and relatively environment-friendly) energy.

\subsection{\label{subsec:subsec12ff}What is the future of quantum computing, quantum information, and other applications of entanglement?}

Quantum computing and quantum information have many facets. The biggest question currently is whether these areas will ever achieve practical importance, because of the fragility of entangled states to decoherence in a realistic environment. The general issue of entanglement is of increasing interest in contexts ranging from physical realizations of quantum computers to resolution of the black hole information paradox.

\subsection{\label{subsec:subsec12gg}What is the future of quantum optics and photonics?}

Photons as well as electrons play an important role in this general context, and in potential new technologies based on photonics, including optoelectronics. The frontiers involve shorter laser pulse durations, higher intensities, radiation at previously inaccessible wavelengths, control of quantum phenomena, and a wide range of emerging new ideas.

What new phenomena will be discovered involving photons alone, or photons acting in concert with electrons and other particles? 


\section{\label{sec:sec4}Deep issues}

Some will consider the following questions to be more metaphysics than physics, but -- given sufficient time and resources -- we believe that they can in principle be addressed by scientists of the (relatively near or very remote) future. An example is the first topic: Higher dimensions might seem as real as atoms if they were equally successful in predicting and explaining experimental data. Similarly, a theory that \textit{postdicts} known physics will be taken seriously if it additionally \textit{predicts} new phenomena which are subsequently confirmed by experiment.

\subsection{\label{subsec:subsec15}Higher dimensions, with geometry and topology of an internal space}

The original idea of Kaluza (and later Klein) was that a purely gravitational theory in 5 dimensions yields gravity plus electromagnetism in 4-dimensional spacetime, if the extra dimension is compactified. Einstein found this idea interesting, even though he took two years to review and accept Kaluza's paper. It was later shown by DeWitt and others that gravity in D dimensions yields gravity plus nonabelian gauge fields in external spacetime, if the extra (D-4) dimensions are compactified~\cite{kaluza1,kaluza2}. Extra dimensions are, of course, currently employed in string theory~\cite{witten,polchinski,beckers}. 
So a quite deep issue is whether there are in fact extra dimensions beyond the familiar four of ordinary spacetime.

If there are extra dimensions, the next question is the structure of the internal space, including its fields. The laws of Nature are presumably determined by this structure, so different internal spaces will correspond to different universes: An internal space is essentially the genome of a universe. 

The most widely known example is string theory, which is so rich in the possibilities for both spatial manifolds and fields that one guess at the number of possible universes is $10^N$ with $N \sim 500$. 

So, if there are in fact higher dimensions, a second deep issue is the structure of the internal space for our universe. 

\subsection{\label{subsec:subsec17}Validity of the multiverse idea and the anthropic principle}

There are many different facets to the multiverse idea:

(1) If all of Nature is described by a path integral over all possibilities, then each of the internal spaces  of  topic \ref{subsec:subsec15} is the foundation for a different universe with different laws.

(2) For a given internal space, there can be many different initial conditions, and again a multiplicity of universes.

(3) Because of the inadequacies of other inflationary models (and alternatives to inflation), many find the ``eternal inflation'' scenario to be quite plausible~\cite{linde2}. In this scenario, since new universes are continuously arising from old,  the number of universes is multiplied yet again.

(4) If we limit attention to only our own universe, inflation implies that it is vastly larger than the local observable universe (which already contains hundeds of billions of galaxies). In fact, it may be infinite, with a flat or open (hyperbolic) geometry. Then our single (global) universe contains an enormous multiplicity of (local) observable universes.

(5) If we further limit attention to only our own observable universe, the Everett (or ``many-worlds'') interpretation of quantum mechanics implies that there are a vast number of different branches in the state vector, with those on our branch unaware of all the others.

It is conceivable that one is forced to at least some steps in this dizzying multiplicity by the requirement of logical consistency, after one delves into the details and finds that the available alternatives lead to inconsistencies. But the multiverse is, of course, quite controversial because it is outside the domain of normal science. 

Equally controversial is the anthropic principle, which (stated simplistically) means that the universe we inhabit must be one that is attuned to the requirements for intelligent life to emerge. There are many versions of this principle, which is attributed to Brandon Carter; his strong anthropic principle states that ``the universe (and hence the fundamental parameters on which it depends) must be such as to admit the creation of observers within it at some stage''. 

This principle is usually motivated by the multiverse concept and the fact that many of the features of our universe appear to be improbably favorable for our existence. We then live in a ``Goldilocks universe'', just as astronomy has shown that we live on a ``Goldilocks planet'' (although we can aspire to avoid the precedent of Goldilocks, who broke her ``just-right'' chair into pieces).

A challenge is to make the multiverse concept and/or anthropic principle part of proper science.

\subsection{\label{subsec:subsec15b}Geometry and topology of external spacetime}

Nonabelian gauge theories predict various kinds of topological defects that might be important in cosmology. These include monopoles, cosmic strings, and domain walls.

In addition, Einstein gravity permits various exotic topologies in spacetime (some of which have made their way into the popular imagination  -- for example, Einstein-Rosen wormholes). Evidence of a nontrivial topology of the universe is occasionally searched for in the CMB measurements.

Also still unresolved are the possibilities of naked singularities and closed timelike loops, which would in principle permit backward time travel.

\subsection{\label{subsec:subsec16}Origin and fate of the universe}

There are many mysteries concerning the origin of the universe. One is why there should have been an origin at all. A second is what fields were initially present, and in what state. Still another is why the initial entropy was so surprisingly low, allowing us to define the future as the direction in which entropy increases. Finally, there is the issue of what –- if anything -- may have preceded the beginning of our particular universe.

\begin{figure}[htbp]
\centering
 \includegraphics[bb=0 0 360 770, width=2in]{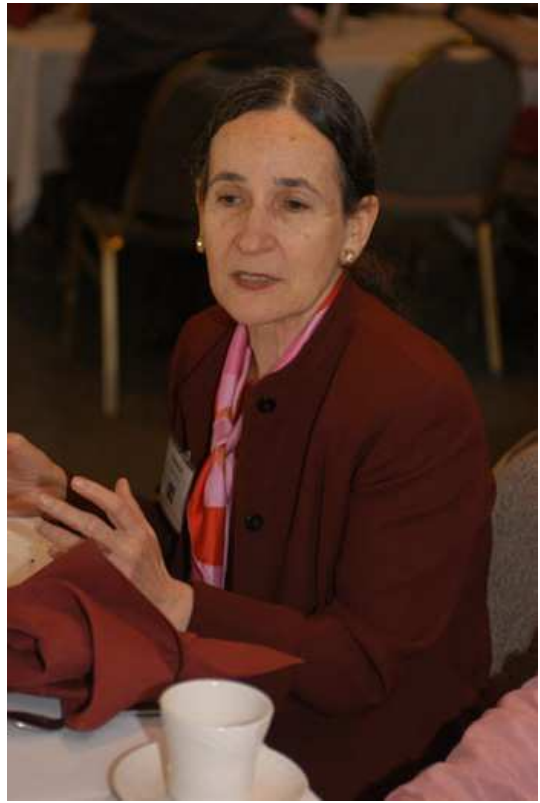}
 \caption{Neta Bahcall, professor at Princeton, played a leading role in developing the ``concordance model'' of modern cosmology. 
\label{Bahcall}}
 \end{figure}

\begin{figure}[htbp]
\centering
 \includegraphics[bb=0 0 340 470, width=2.3in]{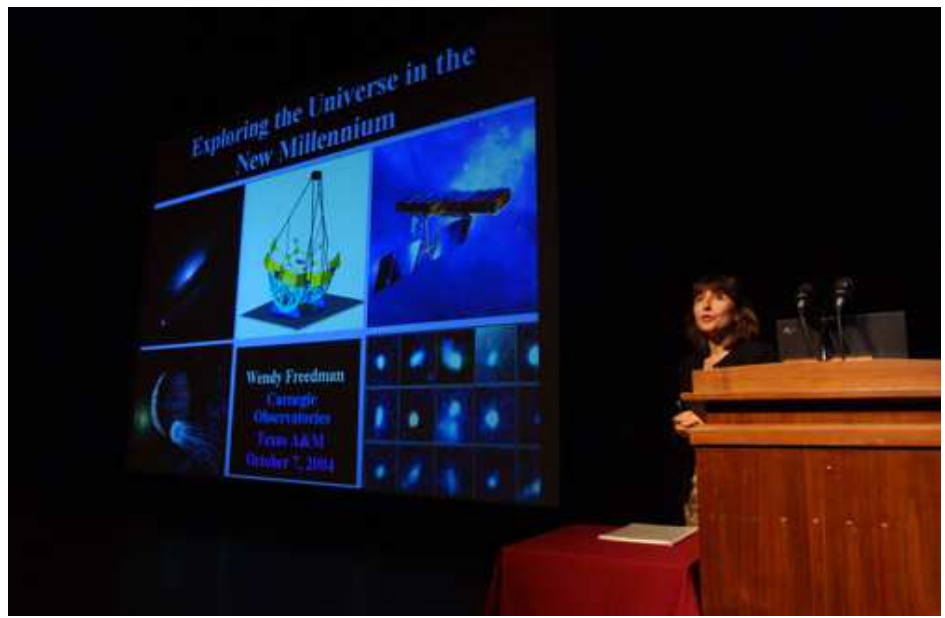}
 \caption{Wendy Freedman, now at the University of Chicago and the Kavli Institute for Cosmological Physics,  gives a public talk on astronomy and cosmology. She led the team that set a new standard for measuring the rate of expansion of the universe, and is currently chair of the board of directors of the Giant Magellan Telescope project.
\label{Freedman}}
 \end{figure}

\begin{figure}[htbp]
\centering
 \includegraphics[bb=0 0 340 470, width=2.3in]{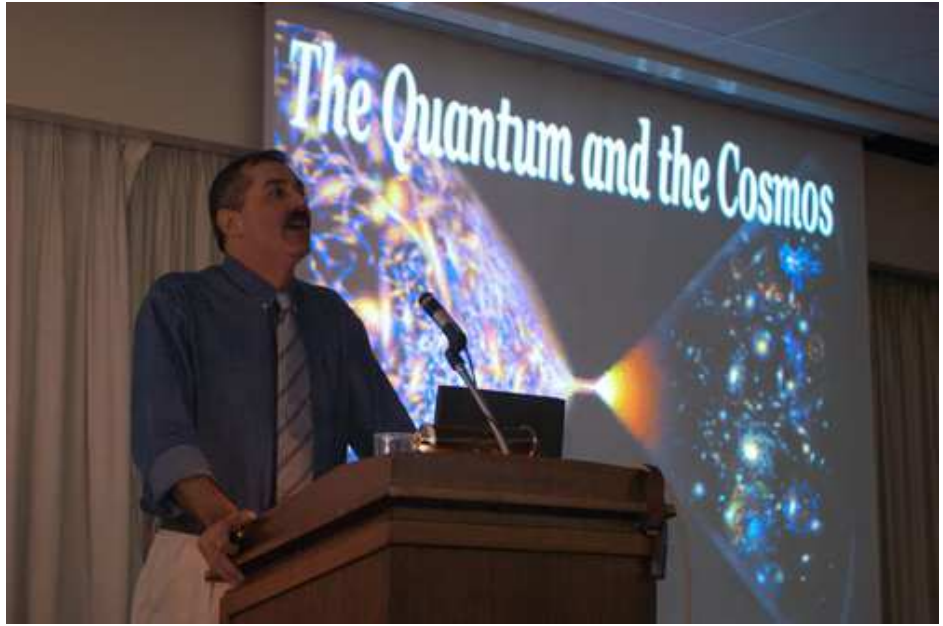}
 \caption{Rocky Kolb of the University of Chicago and Fermilab gives a public talk on inflation and other aspects of cosmology.\label{Kolb}}
 \end{figure}

Figures~\ref{Bahcall}, \ref{Freedman}, and \ref{Kolb} show three of the many astronomers and cosmologists whose work has led to the modern ``concordance'' model of the universe, in which data from various complementary probes are finally found to be consistent.
Since the ``dark energy'' is still being characterized, and it is likely that more cosmological surprises lie ahead, we do not know what course the universe will follow in the future. The most straightforward extrapolation is that cosmic acceleration will cause galaxies to pull away from one another, while each galaxy remains gravitationally bound. More speculative projections postulate a dark energy fluid with a ratio $w$ of pressure to energy density which is different from the value of $-1$ for the vacuum energy -- with e.g. $w < -1$ implying a ``big rip'' that would ultimately pull all matter apart. There are also cyclic models, higher-dimension models, etc. with different implications for both past and future. And, as mentioned in topic \ref{subsec:subsec14}, there are even conjectures that the universe may undergo a phase transition in the remote future to a state with very different properties.
More precise observations will help to exclude many models, but a confident prediction will surely require  fundamental breakthroughs involving gravity, particle physics, and cosmic evolution.

\subsection{\label{subsec:subsec19}What is the origin of spacetime, why is spacetime four-dimensional, and why is time different from space?} 

There have been conjectures that it might be possible to derive spacetime from some more fundamental framework, e.g. in the general context of string theory~\cite{seiberg}, but no convincing treatment has yet been given.

A truly satisfying theory might provide a (nonanthropic) explanation of why the spacetime of ordinary experience has precisely four dimensions. 

Time is distinguished from space only by the signature of the metric tensor. One can formulate a metric theory with 0, 2, ...  time coordinates rather than 1.  What in a fundamental theory might explain why there is exactly one time coordinate? 

\subsection{\label{subsec:subsec21}Origin of Lorentz invariance and Einstein gravity} 

Essentially all attempts at new unified theories -- grand unification, supersymmetry, string theory, ... -- assume local Lorentz invariance (i.e. Einstein relativity), rather than trying to explain it. 

There have been attempts by Sakharov and others to derive gravity from the vacuum energy or some other form of metric elasticity, but none of these efforts have proved convincing. Somewhat in the earlier spirit of Feynman and others, string theory derives gravity as a spin-2 field. But then one has the question of where string theory, its fields, and its action come from, and the suggestions in this direction -- like string theory itself -- have not met with wide acceptance. 
\begin{figure}[htbp]
\centering
 \includegraphics[bb=0 0 340 450, width=2.8in]{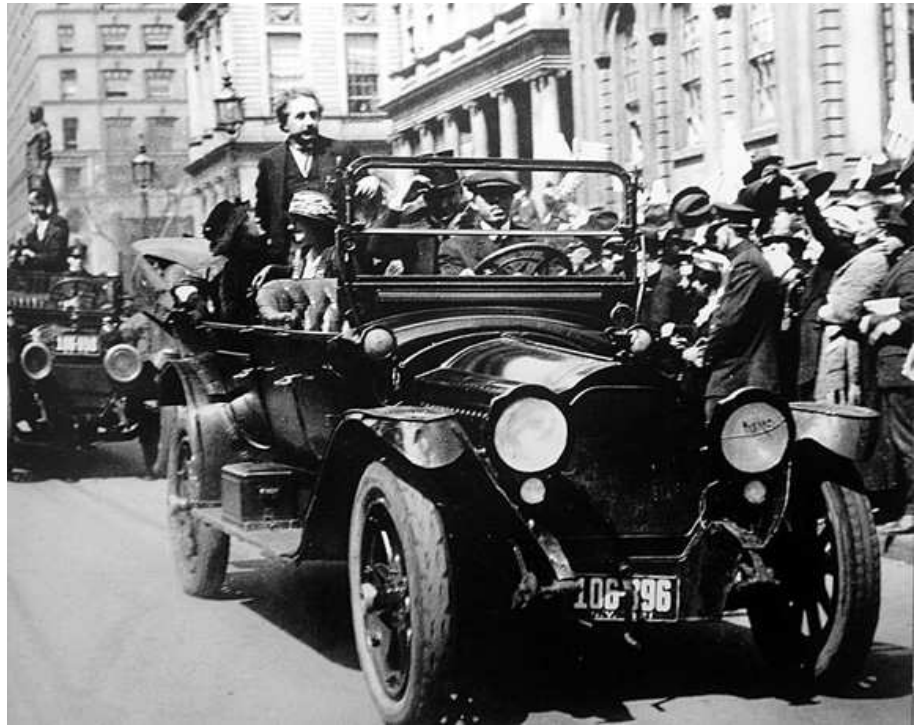}
 \caption{Albert Einstein, in New York in 1921, had been transformed from a modest scientist to a world-wide celebrity. His 1916 prediction of gravitational waves has recently been verified, but his even earlier concerns about the interpretation of  quantum mechanics have not yet been satisfactorily answered.
Credit: Wikimedia Commons and Life magazine.
 \label{Einstein}}
 \end{figure}
So the fundamental origin of gravity also remains an open issue. 

\subsection{\label{subsec:subsec20}Origin of gauge fields, their coupling to matter fields, and their action}

All the forces of the Standard Model are described by gauge fields. (Even gravity is described by a gauge theory, although with a different structure.) A truly fundamental theory might explain why Nature has chosen only these kinds of forces. It might also explain why matter has a simple minimal coupling to these fields, and why their action has such a simple minimal form. 

\subsection{\label{subsec:subsec23}Origin and interpretation of quantum mechanics and quantum fields}

A truly fundamental theory might derive quantum mechanics and quantum fields from a deeper principle.

There is also the issue of the interpretation of quantum mechanics, on which there is still no universal agreement. In 1911 Einstein, shown in Fig.~\ref{Einstein},  already recognized the difficulty of this problem. Despite the vast number of articles and books written on this subject, his criticisms have never been satisfactorily answered, in the sense that most knowledgeable physicists are still puzzled by wave-particle duality (or the Schr\"{o}dinger's cat and Einstein-Podolsky-Rosen paradoxes, which similarly arise in the picture with wavefunction collapse during a measurement or observation). 

Regardless of interpretation, the deepest imaginable fundamental theory would explain why we live in a universe which consists of quantum fields, and how those fields originate.

\subsection{\label{subsec:subsec25d}Mathematical consistency} 

A theory should be mathematically, logically, and philosophically consistent, as well as consistent with experiment and observation. But the mathematical consistency of even simple quantum field theories in four spacetime dimensions has not yet been rigorously established. 

\subsection{\label{subsec:subsec27}Connection between the formalism of physics and the reality of human experience} 

The words of Stephen Hawking~\cite{brief} reflect the point of view of a mathematical physicist: ``What is it that breathes fire into the equations ...." One might reverse this point of view by starting with Nature, whose basic principles we are far from understanding, and recognizing that mathematical physics is essentially a  human creation that bears the same relation to Nature itself as a map bears to the rich terrain that it schematically depicts. 

One might hope that the most fundamental theory will eventually reveal the essential character of reality (Kant's  ``Ding an sich'') which now is directly known to us only in one way -- through the experiences of our conscious minds.

Under this general topic -- the ultimate nature of reality -- we subsume the old question ``Why is there something rather than nothing?'' The simplest answer is ``Why not?'', and the most amusing comes from Sidney Morgenbesser: ``If there were nothing you'd still be complaining!''

But the best current idea has been stated by Frank Wilczek: ``The answer to the ancient question `Why is there something rather than nothing?' would then be that `nothing' is unstable.'' This idea is treated more expansively in a book by Lawrence Krauss~\cite{Krauss}.

A quotation from Einstein concerns an issue that may be equally deep: ``The most incomprehensible thing about the world is that it is comprehensible.'' What principle explains the fact that the present universe evolves smoothly according to simple laws, and is not instead a random chaotic mess?

One expects that an ultimate understanding of Nature will authenticate the Emily Dickinson lines:
\begin{center}

Nature is what we know	

But have no art to say,	 

So impotent our wisdom is
	
To Her simplicity.

\end{center}

\section{\label{sec:sec5}Potential for breakthroughs in techniques and technology}

\begin{figure}[htbp]
\centering
 \includegraphics[bb=0 0 160 310, width=2.6in]{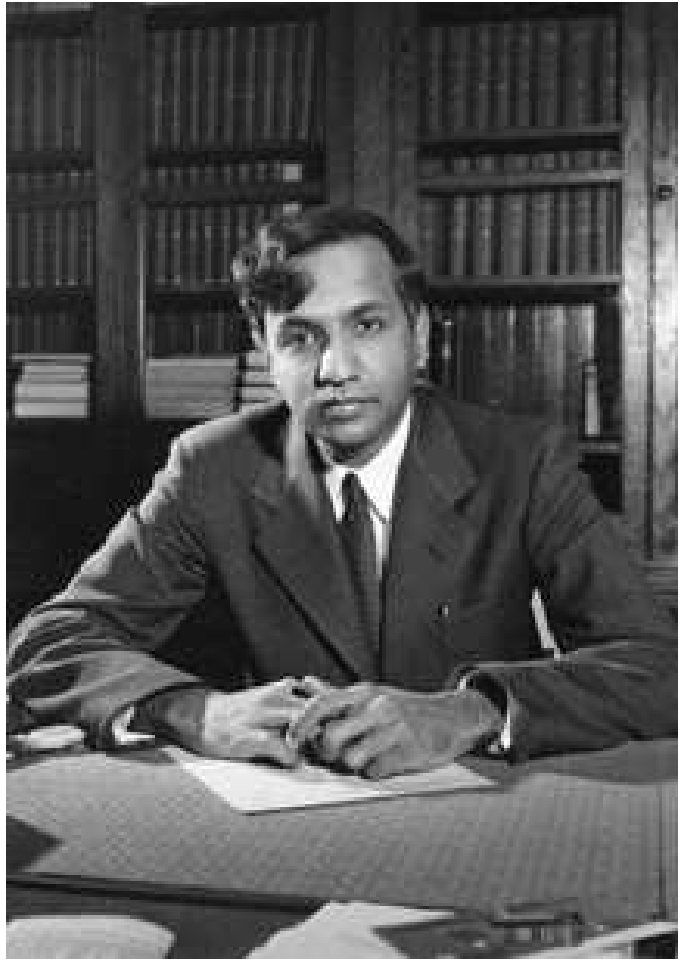}
 \caption{Subrahmanyan Chandrasekhar is considered by many to be the greatest of theoretical astrophysicists. His most famous result is the Chandrasekhar limit for the mass of white dwarf, with a value of approximately 1.4 times the mass of the Sun. When a white dwarf acquires more than this mass -- for example, by accreting mass from another star in a binary system, or by merging with another star --  there is gravitational collapse followed by a type Ia supenova. Credit: Photograph by Stephen Lewellyn, Chicago, IL, courtesy AIP Emilio Segre Visual Archives, Tenn Collection.
\label{Chandra}}
 \end{figure}
\begin{figure}[htbp]
\centering
 \includegraphics[bb=0 0 400 800, width=2in]{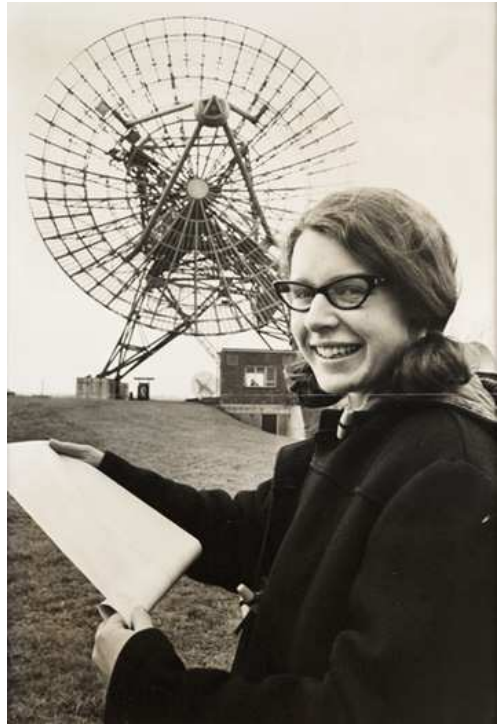}
 \caption{Jocelyn Bell Burnell, discoverer of pulsars. See also Fig.~\ref{Bell-telescope}. Credit: Daily Herald Archive/National Media Museum/SSPL.
\label{Bell}}
 \end{figure}
\begin{figure}[htbp]
\centering
 \includegraphics[bb=0 0 170 250, width=3.2in]{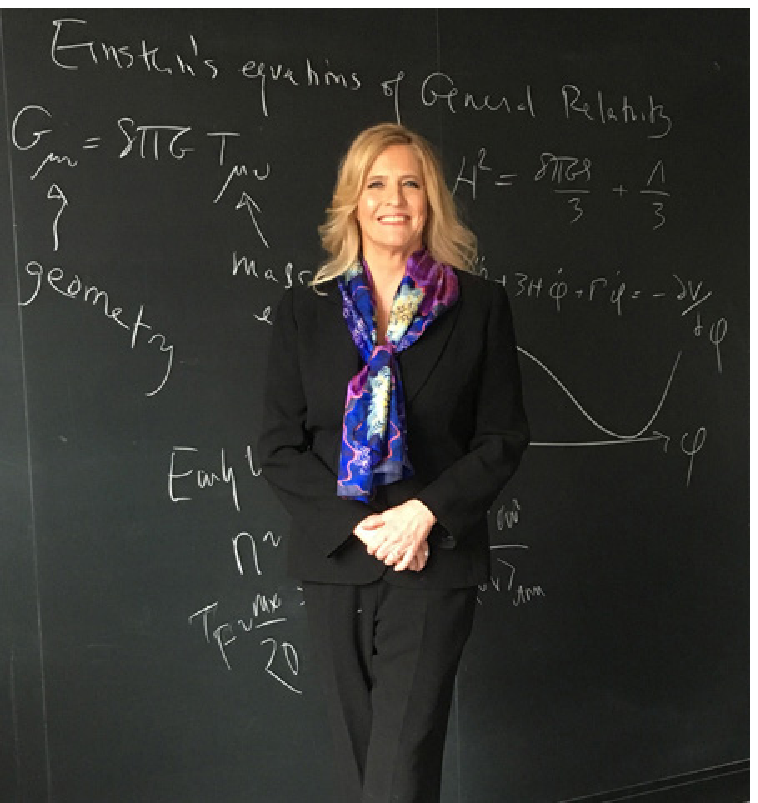}
 \caption{Katherine Freese is George E. Uhlenbeck Professor of Physics at the University of Michigan, Guest Professor at Stockholm University, Director Emerita at Nordita (the Nordic Institute for Theoretical Physics), and author of one of the best popular books on astrophysics, \textit{The Cosmic Cocktail}~\cite{freese}. Her many contributions to theoretical cosmology include the idea of dark stars, powered by the annihilation of dark matter particles before the first ordinary stars were born.  Credit: Katherine Freese.
\label{Freese}}
 \end{figure}

\subsection{\label{subsec:subsec28}Potential for breakthroughs in theoretical, computational, experimental, and observational techniques}

\textit{Theory}:
\bigskip

Most calculations in high energy physics are based on perturbative methods, as exemplified by expansions in terms represented by Feynman diagrams. For example, the current state of the art in perturbative QCD calculations is next-to-next-to-leading-order. It is highly nontrivial to perform these calculations, and to  check their convergence and accuracy.

Existing nonperturbative techniques for realistic calculations are primarily numerical, with the best known  method being lattice gauge theory. But essentially all numerical methods for real systems lead to rapidly increasing demands for computer time and memory, and again it is not trivial to be sure of the convergence and accuracy.

A major breakthrough would be to discover nonperturbative techniques that allow reliably accurate calculation of important properties and processes for real systems, with a paradigm being QCD at arbitrary energy.  One dream in this direction is dualities which would map strong coupling for realistic physical systems into weak coupling for dual systems, analogous to those used for simple models in string theory and condensed matter physics.

This is one example of the potential breakthroughs in theoretical techniques that we can hope will carry us to improved understanding.

\bigskip
\textit{Computation}:
\bigskip

Computation is rapidly becoming a third leg of physics (theory -- computation -- experiment), and breakthroughs in each of the three areas are of equal importance. Realistic simulations are also  becoming increasingly important in technology.

The events recorded by detectors in collider experiments are mind-boggling in both  complexity and quantity. A major breakthrough would be the development of ``intelligent'' software that analyzes events on a nanosecond time scale at the detector and essentially eliminates the possibility of discovery events being discarded. Both the theoretical calculations and the computer science aspects (for analyzing data from a collider detector) are at the outer frontier of software development.

Important phenomena in astrophysics often defy realistic simulation because of the prohibitively large number of degrees of freedom, and it may be that radical computational innovations are required in this context. 

In the much broader arena of other fields of science and technology (and the still larger realm of human affairs), there is a compelling need for computer science to produce increasingly more elegant and more powerful algorithms. 

\bigskip
\textit{Experiment}:
\bigskip

The progression toward higher energies appears to demand major innovations~\cite{aleksan} to achieve facilities like a linear $\sim 0.5$ TeV $e^+ e^- $ collider, a muon collider, a photon collider, or a very large hadron collider, perhaps ultimately taking proton collisions to $\sim 100$ TeV and the much cleaner lepton-antilepton collisions to $\sim 3$ TeV.

The intense environments at the LHC and future colliders present new demands and new opportunities for detector technologies~\cite{cattai}.
\begin{figure}[htbp]
\centering
 \includegraphics[bb=0 0 360 520, width=2.5in]{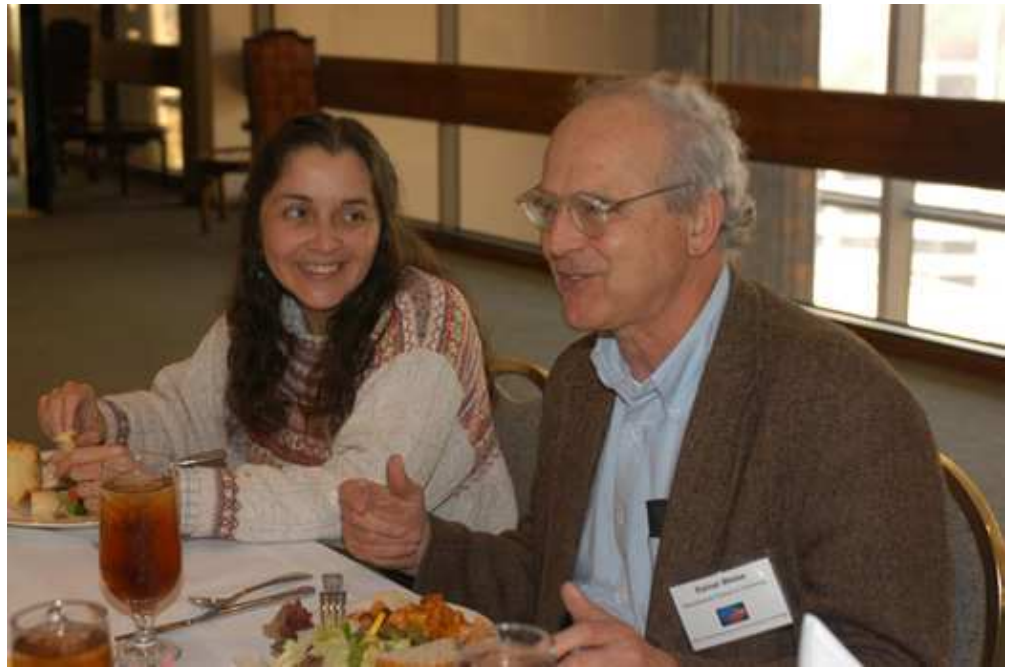}
 \caption{Rainer Weiss led the way to devising the sophisticated laser interferometric techniques which are the basis for LIGO (the Laser Interferometer Gravitational-Wave Observatory). He was also 
the chairman of the science working group for the COBE (Cosmic Background Explorer) satellite mission, which observed the inhomogeneities in the cosmic background radiation that seeded initial structure formation in the early universe, as well as an originator of COBE with John Mather. He has thus been at the center of two of the greatest discoveries in modern physics and astronomy.
Here he is shown with astronomer Sarah Higdon.
\label{Weiss}}
 \end{figure}
Other fundamental experiments (direct dark matter detection, the search for neutrinoless double beta decay, neutrino physics, ...) will employ increasingly large systems, but would also benefit from further innovations, since in some cases they may require well over an order of magnitude increase in sensitivity.
\begin{figure}[htbp]
\centering
 \includegraphics[bb=0 0 360 550, width=2.5in]{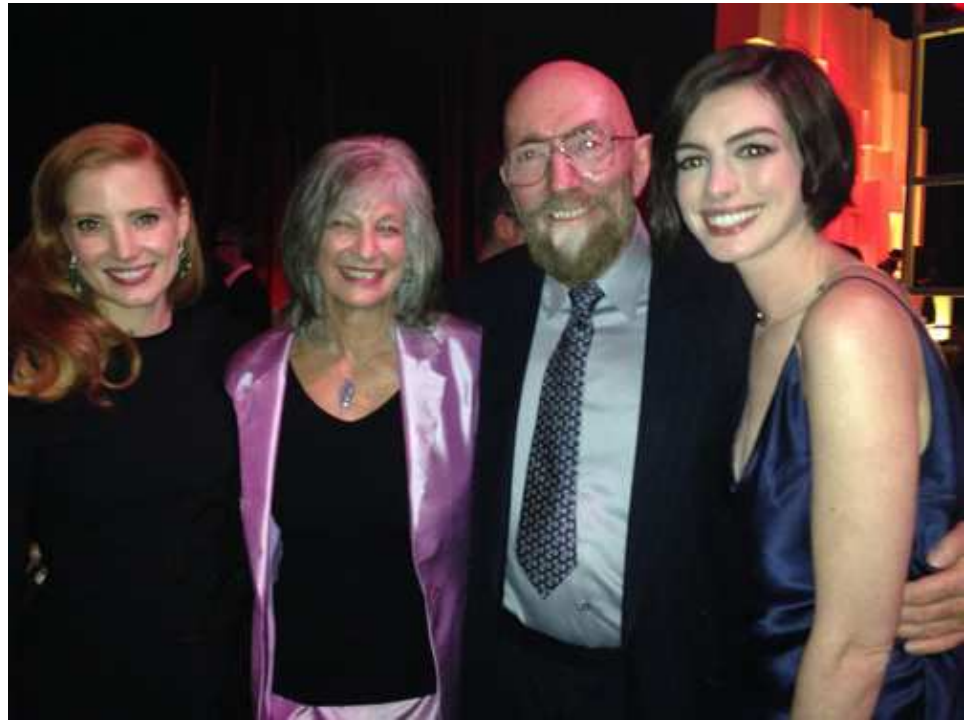}
 \caption{Kip Thorne has also been a principal driving force on LIGO, following a 1975 all-night discussion with Rai Weiss, when they shared a room at a NASA meeting on cosmology and relativity. (Weiss, an experimentalist, had reserved a room, and Thorne, a theorist, had not.) Here Kip Thorne is shown with his wife Carolee Winstein and the stars Jessica Chastain and Anne Hathaway of \textit{Interstellar}. 
Credit: Kip Thorne.
\label{Thorne3}}
 \end{figure}

\begin{figure}[htbp]
\centering
 \includegraphics[bb=0 0 360 340, width=4.5in]{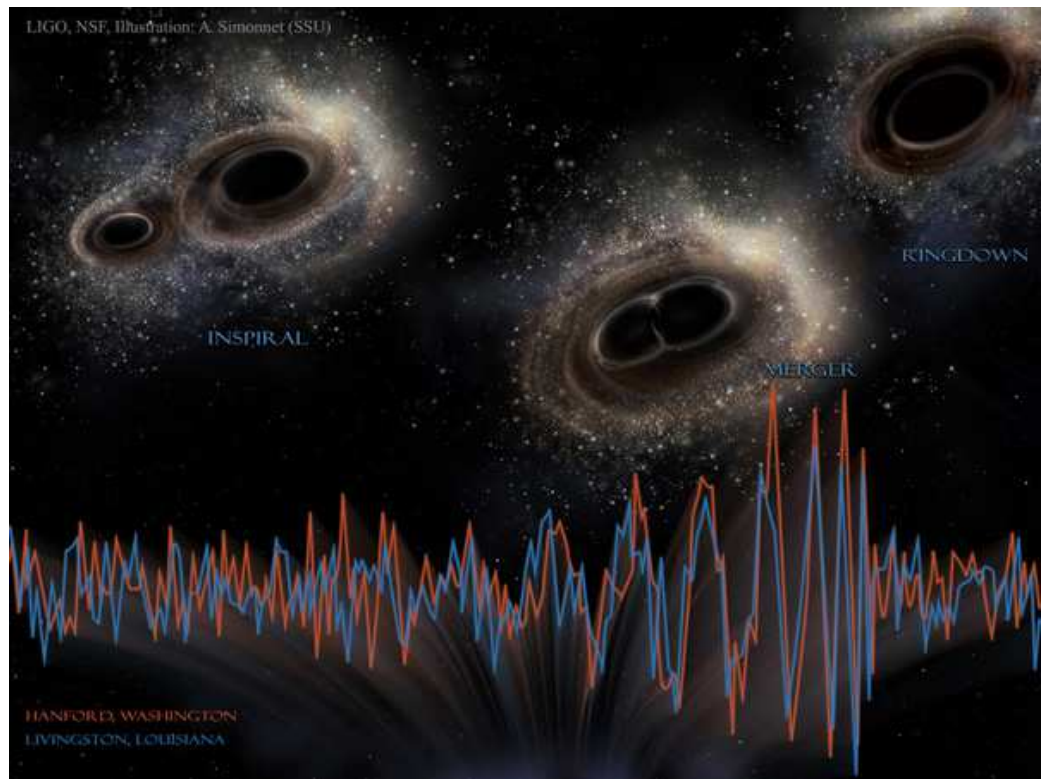}
 \caption{The LIGO collaboration has a spectacular double success: the first direct observation of gravitational waves -- predicted by Einstein a century earlier -- and the unexpected observation of the merger of two very massive black holes.
As indicated in the figure, the waves were simultaneously detected at the two separate LIGO observatories in Hanover, Washington and Livingston, Louisiana, with amazingly consistent signals for the spiralling in and merger of the black holes, and the ringdown of the single remnant afterward. This observation marks the beginning of gravitational wave astronomy.
Credit: LIGO, NSF, Aurore Simonnet (Sonoma State U.).
\label{blackholes}}
 \end{figure}

\bigskip
\textit{Observation}:
\bigskip

Astronomy probes a wide spectrum of exotic phenomena that are inaccessible in terrestrial experiments, using many components of the electromagnetic spectrum plus cosmic rays, neutrinos, and now even  gravitational waves. Figures~\ref{Bell}, \ref{Weiss}, \ref{Thorne3}, and \ref{blackholes} represent the triumphs of radio astronomy (used by Jocelyn Bell Burnell in her discovery of pulsars) and gravitational wave astronomy (pioneered by Rainer Weiss, Kip Thorne, Ronald Drever, Barry Barish, their many collaborators, and others). This last capability, in particular, opens a promising new window on the universe, and has already demonstrated the unexpected importance of intermediate-mass black holes.

There are still many mysteries in astrophysics, which may be resolved with yet more advanced technology and more sophisticated methods. 

\subsection{\label{subsec:subsec77}The ultimate limits of chemistry, applied physics, and technology}

The variety of substances created by inorganic processes (e.g. in geology) is remarkable, while the number of those exploited in biological systems is much larger still. And there appears to be no limit to the complexity of chemical systems that we ourselves can design for future applications. Similarly, there is an ever-increasing richness of possibilities in areas like condensed matter physics and quantum optics. If one extends the discoveries of the past two centuries into the next million or billion years, what technologies may completely transform the lives of our descendants?

One technology is artificial intelligence, which may be computer-like (based on classical bits) or human-like (based on neuronal connections) or something else entirely (based e.g. on quantum states, as in qubits or qudits). How will our descendants make use of the full range of emerging technologies and an evolution that they themselves control?

\section{\label{sec:sec6}Life}

\subsection{\label{subsec:subsec100a}What is life?}

In 1944 Erwin Schr\"{o}dinger wrote a small book with this title, containing the comment that ``We have inherited from our forefathers the keen longing for unified, all-embracing knowledge.'' This statement may be interpreted as an explanation of why physicists have the hubris to discuss problems on which they are not experts. 

Viruses are at the border between living and nonliving systems, since they cannot replicate on their own, but are very efficient at propagating if they can employ normal living cells. This fact was known before Schr\"{o}dinger wrote his book, but today his question may have an even broader context than it did 70 years ago. What forms of life might be based on exotic biochemistry, perhaps not employing DNA as the central molecular structure, or even carbon as the central element? Perhaps currently unknown principles could give rise to unfamiliar lifeforms  on the exoplanets that are now known to be abundant, or even in different universes if the multiverse idea has any validity. 

After an extremely creative and successful early career in astronomy, and after being knighted, Fred Hoyle turned to writing popular science books and science fiction novels. One of these was \textit{The Black Cloud} (published in 1957) in which an immense cloud of gas entering the solar system turns out to be sentient, much more intelligent than human beings, and surprised to find intelligent life on the surface of a planet. This was a precursor of other ideas exploring the possibilities for life far outside the boundaries of ordinary experience. 

Despite many fictional and serious considerations of this topic, there is still no generally accepted answer to the simple question in the title of this subsection.

\subsection{\label{subsec:subsec100b}How did life on Earth begin -- and how did complex life originate?}

Earth was formed in the early Solar System about 4.5 billion years ago, and the earliest accepted evidence for life dates back about 3.5 billion years. These facts and others suggest that life on Earth has passed through two phases, each lasting roughly 2 billion years -- the first with only one-celled prokaryotes, and the second leading up to a remarkable proliferation of multicelled eukaryotes~\cite{fortey}. Even a single bacterium, archaeon, or eukaryotic cell deserves respect for its multiple functioning components, but the complexity of e.g. a human body is truly amazing.
\begin{figure}[htbp]
\centering
 \includegraphics[bb=0 0 360 840, width=1.5in]{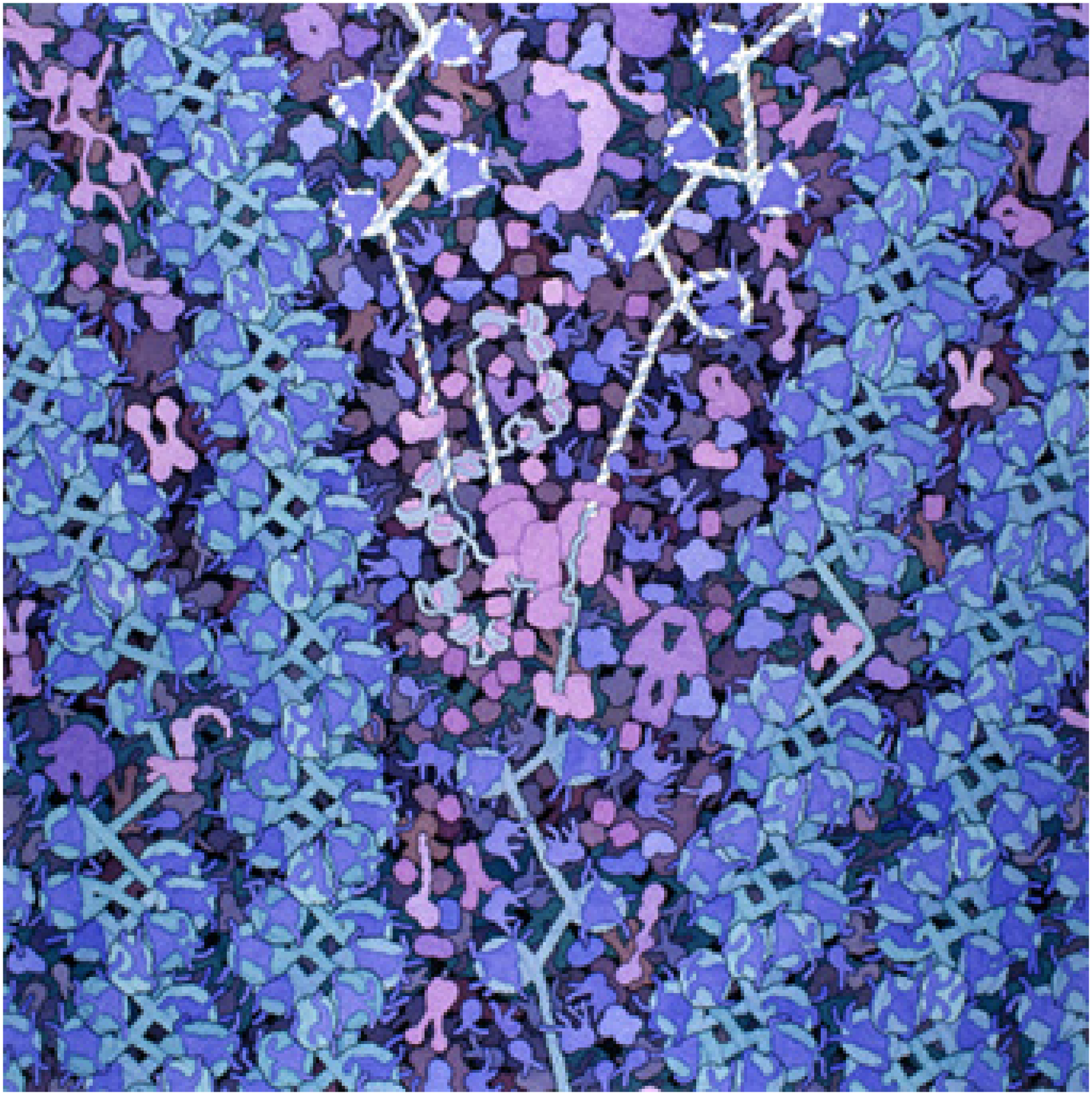}
 \caption{Depiction of  DNA polymerase (molecule in center) replicating DNA, with chromatin on either side. Credit: David Goodsell.
\label{biosites}}
 \end{figure}

There are various theories of how life arose on Earth, and currently none are fully convincing. Perhaps the biggest question is whether life developed afresh from organic molecules on an early Earth, or instead descended to Earth (in primitive form) after first arising elsewhere~\cite{universe}. Experiments and genetic analyses strongly suggest that the last universal common ancestor (LUCA) lived near hot deep sea vents, where seawater interacts with magma escaping through the ocean floor~\cite{LUCA}. Since all life forms on Earth have evolved out of this remote ancestor, they share some common attributes and molecules, such as DNA, depicted as being replicated in Fig.~\ref{biosites}.
\begin{figure}[htbp]
\centering
 \includegraphics[bb=0 0 360 540, width=4in]{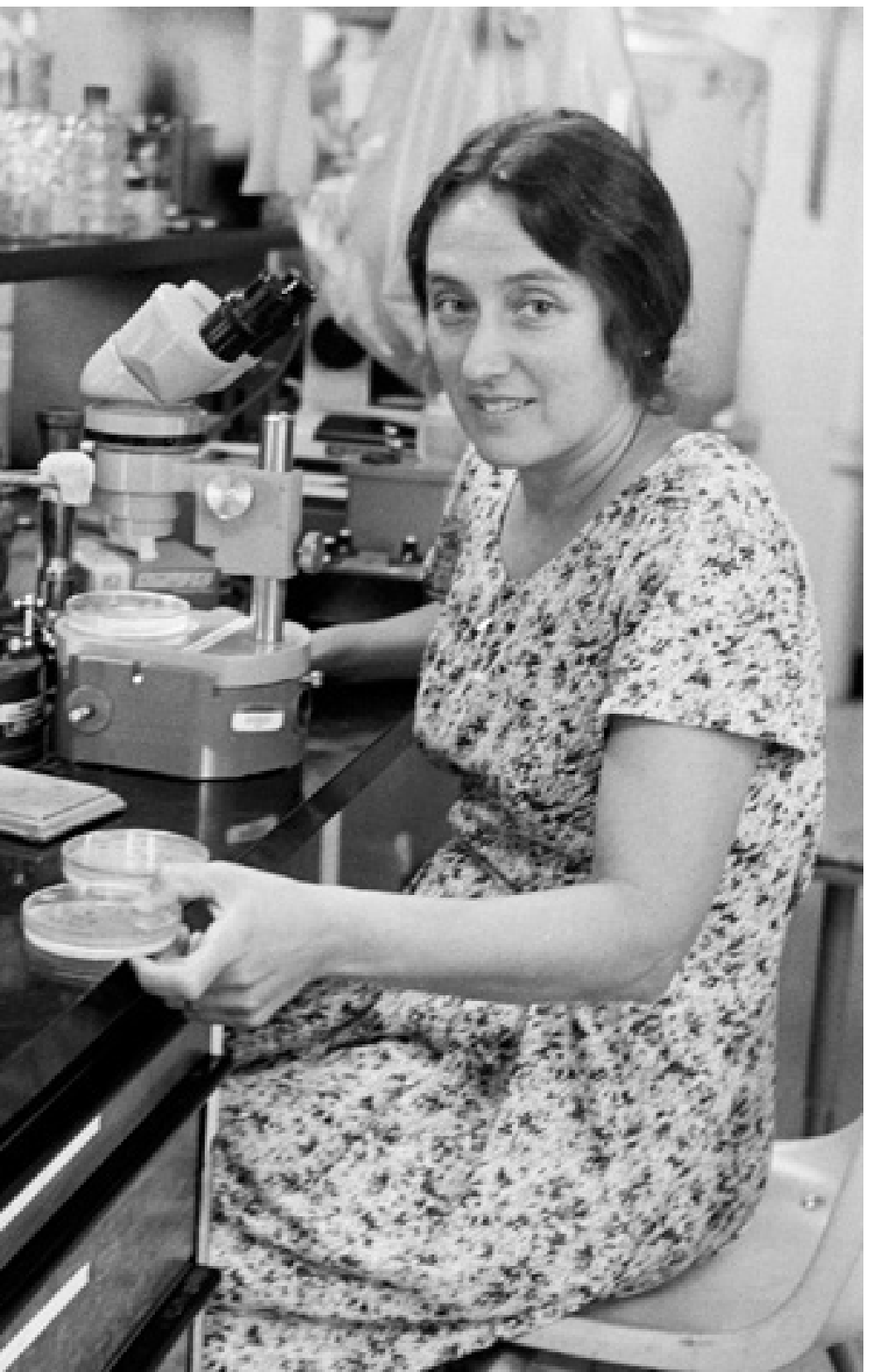}
 \caption{Lynn Margulis courageously proposed a theory which was once ignored (or derided) but is now generally accepted, after decisive experimental confirmation. It is  of central importance for understanding the origin of multicelled animals, plants, and fungi, as well as other eukaryotic organisms -- all the familiar life that is based on cells with nuclei. Her basic idea was that cell organelles such as mitochondria and chloroplasts were once independent bacteria. After years of initial rejection, she finally received major recognition and honors, including the American National Medal of Science from President Bill Clinton. Credit: BU (Boston University) Photography.
\label{margulis}}
 \end{figure}

A separate and equally important question is how \textit{complex} life arose from its one-celled precursors. A basic idea due to Lynn Margulis, shown in Fig.~\ref{margulis}, is now accepted: The  mitochondria and chloroplasts in eukaryotic cells were once independent bacteria. According to one interpretation~\cite{lane}, life would always have been limited to one-celled bacteria and archaea (prokaryotes) had not an archaeon undergone a symbiotic merger with a bacterium which ultimately led to the last of the three domains of life on earth -- the eukaryotes. 

\subsection{\label{subsec:subsec100c}How abundant is life in the universe, and what is the destiny of life?}

\begin{figure}[htbp]
\centering
 \includegraphics[bb=0 0 160 330, width=2.4in]{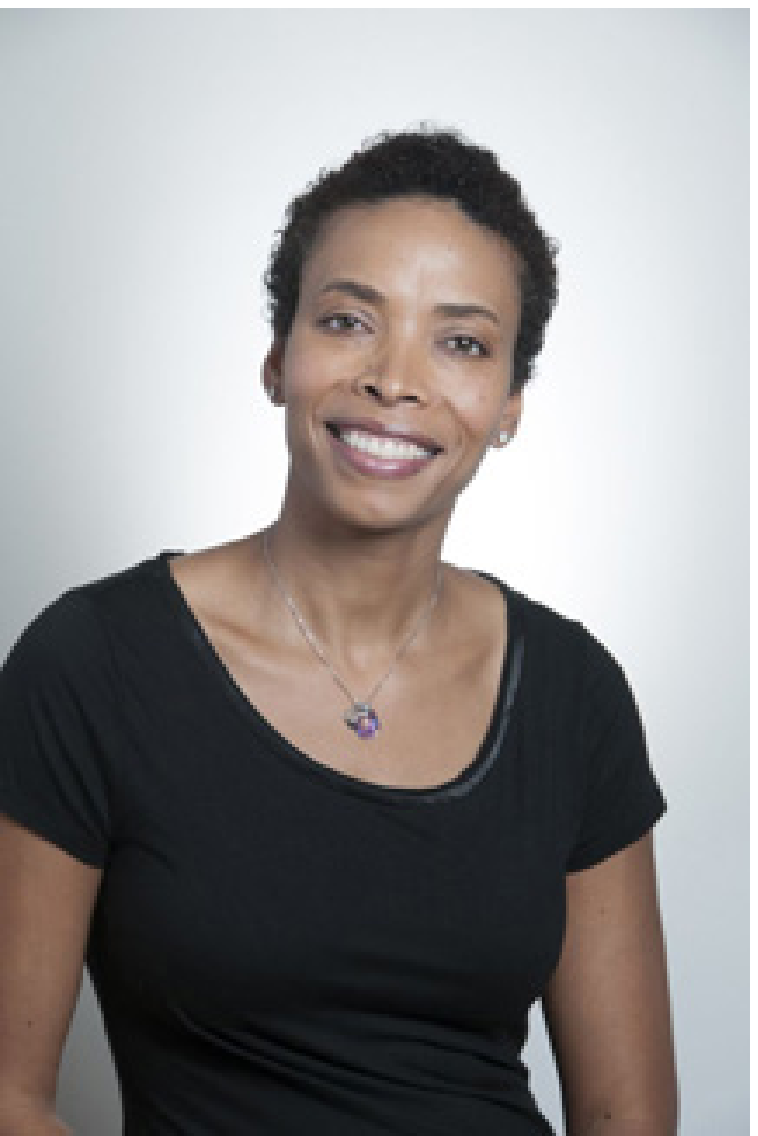}
 \caption{Aomawa Shields is a National  Science Foundation Postdoctoral Fellow in Astronomy and Astrophysics, and a University of California President's Postdoctoral Fellow in the UCLA Department of Physics and Astronomy and the Harvard-Smithsonian Center for Astrophysics. Among other activities, she is Founder and Director of the Rising Stargirls Workshop, with one topic being the search for life elsewhere in the universe. Credit: Martin Cox, http://newsroom.ucla.edu/releases/planet-1-200-light-years-away-good-prospect-for-habitable-world.
\label{Shields}}
 \end{figure}

During the past two decades, thousands of exoplanets have been discovered, with a few having characteristics that may be hospitable for life. This fact suggests that life should be quite abundant, in either of the two scenarios for the origin of life mentioned above, given the billions of galaxies in the observable universe and the billions of stars per galaxy.

If one additionally considers the billions of years that elapsed even before our own Solar System was born, providing much earlier opportunities for intelligent life to evolve, one is led to Enrico Fermi's question regarding other civilizations with advanced technologies: Where are they? This question has long been addressed by many astronomers, including Jill Tarter (former director of the Center for SETI Research, and principal inspiration for the astronomer portrayed by Jodie Foster in the movie ``Contact'') and Aomawa Shields, shown in Fig.~\ref{Shields}.

There are many imaginable answers, including the possibilities that super-intelligent life avoids contact with lesser civilizations or destroys itself through increasingly hazardous technologies. But it may be that intelligent life is just extremely rare, because of the many bottlenecks to its development~\cite{Human}.

Turning to the future, it is beginning to appear that we will soon be able to control the genetic endowment and attributes of those yet to be born. Controversies have already arisen over how this capability should be used in the near future. On a longer time scale, of a million or perhaps a billion years, there are still more profound questions about the legacy we wish to leave.

And if there is intelligent life elsewhere, these beings will be making their own decisions.

\subsection{\label{subsec:subsec100x}How does life solve problems of seemingly impossible complexity?}

There are profound mysteries as to how organisms are able to achieve at least two feats of seemingly impossible complexity, which are far beyond the ability of current computer simulations to replicate.
The first is protein folding, in which a protein chain forms up into its correct biologically functional structure. The second is morphogenesis: As an initial single cell multiplies into the complete organism, signals that are currently not understood tell the differentiating cells to form up into intricate structures like eyes, heart, brain, and other organs. 

\subsection{\label{subsec:subsec100y}Can we understand and cure the diseases that afflict life?}

The biological pathways in any organism are bewilderingly complex, although many have been mapped out. Given the vast number of degrees of freedom -- and the fact that e.g. no two human bodies are identical -- to what extent is it possible to understand the origins of diseases, and how they can be prevented or cured? Is this a strictly empirical enterprise, or can theoretical systems biology make a substantial contribution?

We include aging (and longevity) within this general topic.

\subsection{\label{subsec:subsec100d}What is consciousness?}

Our only direct contact with reality is through our own experiences, which science places in the neuronal structures of the brain, and which are now being probed with the tools of neuroscience. Various mental processes have been revealed to take place in specific areas, but it is still not established what physical processes correlate with consciousness --  i.e., our global awareness of the input from our senses and our internal mental processes. A primary issue is whether consciousness is localized in one region or is instead distributed throughout the brain.

Another major issue is what physical systems will support the real experiences that we associate with consciousness, and how can we tell whether another being has such experiences? A normal Turing test is not sufficient!

\section{\label{sec:sec7}Who will solve the biggest problems?}

\begin{figure}[htbp]
\centering
 \includegraphics[bb=0 0 1721 3001, width=4.5in]{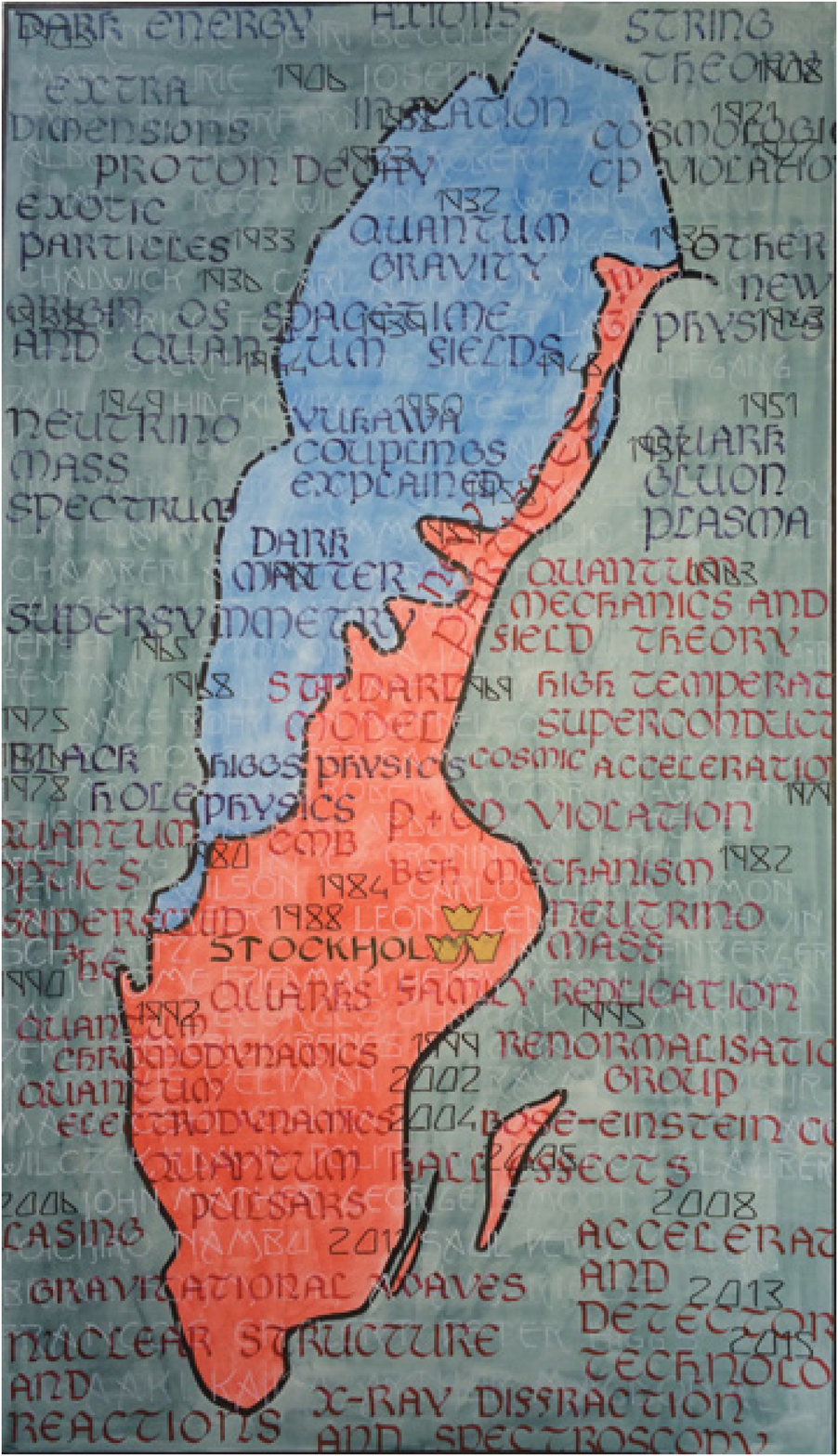}
 \caption{Coming in from the cold. The black line symbolically separates the charted territory in our current worldview from the territory still under exploration. These regions are respectively shaded red and blue, with the blue topics emphasized in this article. In the background are previous Nobel Laureates in fundamental physics.\label{cold}}
 \end{figure}

The Nobel Prizes in physics, chemistry, and medicine and physiology have long served to provide definitive recognition for the greatest achievements in science. In Fig.~\ref{cold}, we therefore show a map of Sweden, together with some representative achievements of the past, in red, and potential achievements of the future, in blue. The boundary between the warmer and colder regions -- specifically, the 1000 degree-day line -- symbolizes the boundary between what is known and what is currently being explored. The northern and more mountainous region can be an exhilarating place to inhabit, full of natural wonders, but the road to Stockholm may also prove quite rewarding~\cite{road}. It is likely that some of those pursuing the topics discussed above will in fact eventually make their way to Stockholm to enjoy the cozy warmth of established physics.

\bigskip\noindent
In the inspiration for this article -- \textit{The Hitchhiker's Guide to the Galaxy} -- the Ultimate Question itself is left undetermined after many adventures, and only the answer is known: ``Forty-two". Our interpretation of the Ultimate Question is ``How many fundamental issues must be resolved if we are to attain full enlightenment about Life, the Universe, and Everything?" and our first draft (or personal selection) can be summarized in the form of simplified questions:

\bigskip\noindent
1. Why does conventional physics predict a cosmological constant that is vastly too large?

\noindent
2. What is the dark energy?

\noindent
3. How can Einstein gravity be reconciled with quantum mechanics?

\noindent
4. What is the origin of the entropy and temperature of black holes?

\noindent
5. Is information lost in a black hole?

\noindent
6. Did the universe pass through a period of inflation, and if so how and why?

\noindent
7. Why does matter still exist?

\noindent
8. What is the dark matter?

\noindent
9. Why are the particles of ordinary matter copied twice at higher energy?

\noindent
10. What is the origin of particle masses, and what kind of masses do neutrinos have?

\noindent
11. Does supersymmetry exist, and why are the energies of observed particles so small compared to the most fundamental (Planck) energy scale?

\noindent
12. What is the fundamental grand unified theory of forces, and why?

\noindent
13. Are Einstein relativity and standard field theory always valid?

\noindent
14. Is our universe stable?

\noindent
15. Are quarks always confined inside the particles that they compose?

\noindent
16. What are the complete phase diagrams for systems with nontrivial forces, such as the strong nuclear force?

\noindent
17. What new particles remain to be discovered?

\noindent
18. What new astrophysical objects are awaiting discovery?

\noindent
19. What new forms of superconductivity and superfluidity remain to be discovered?

\noindent
21. What new topological phases remain to be discovered?

\noindent
20. What further properties remain to be discovered in highly correlated electronic materials?

\noindent
22. What other new phases and forms of matter remain to be discovered?

\noindent
23. What is the future of quantum computing, quantum information, and other applications of entanglement?

\noindent
24. What is the future of quantum optics and photonics?

\noindent
25. Are there higher dimensions, and if there is an internal space, what is its geometry?

\noindent
26. Is there a multiverse?

\noindent
27. Are there exotic features in the  geometry of spacetime, perhaps including those which could permit time travel?

\noindent
28. How did the universe originate, and what is its fate?

\noindent
29. What is the origin of spacetime, why is spacetime four-dimensional, and why is time different from space?

\noindent
30. What explains relativity and Einstein gravity? 

\noindent
31. Why do all forces have the form of gauge theories?

\noindent
32. Why is Nature described by quantum fields?

\noindent
33. Is physics mathematically consistent?

\noindent
34. What is the connection between the formalism of physics and the reality of human experience?

\noindent
35. What are the ultimate limits to theoretical, computational, experimental, and observational techniques?

\noindent
36. What are the ultimate limits of chemistry, applied physics, and technology?

\noindent
37. What is life?

\noindent
38. How did life on Earth begin -- and how did complex life originate?

\noindent
39. How abundant is life in the universe, and what is the destiny of life?

\noindent
40. How does life solve problems of seemingly impossible complexity?

\noindent
41. Can we understand and cure the diseases that afflict life?

\noindent
42. What is consciousness?

\begin{figure}[htbp]
\centering
 \includegraphics[bb=0 0 380 680, width=1.9in]{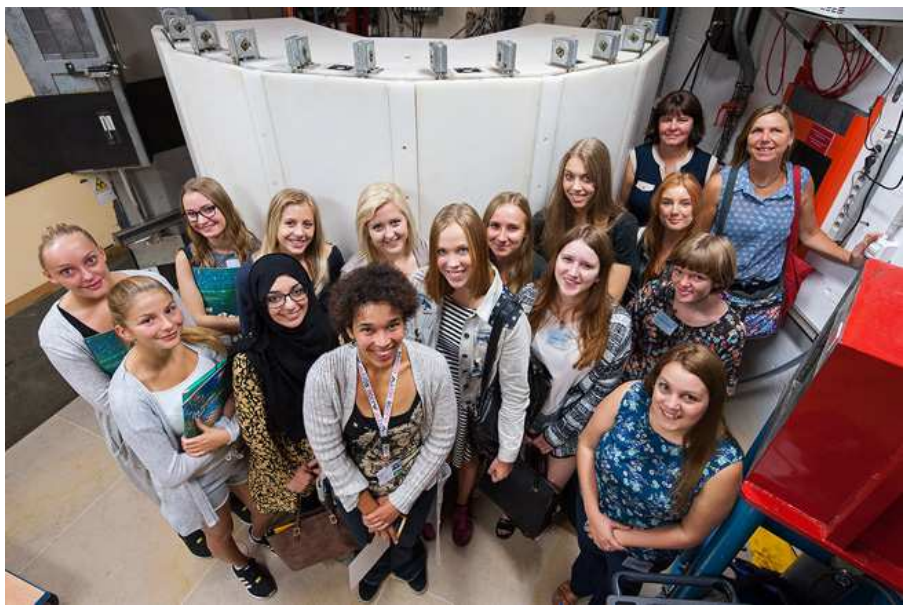}
 \caption{Twelve prizewinning Swedish school girls, with outreach organisers from Uppsala and Ume\r{a} Universities in Sweden and Warwick University, UK, on a visit to the Institut Laue-Langevin in Grenoble. Their guide, centre front, was from Gabon. The girls' parents were born in Finland, Iran, Iraq, Poland and Sweden. During their visit they met or were accompanied by English, French, Icelandic, Italian, Russian, and Swedish scientists. Photograph by Max Alexander. Copyright: Uppsala University.
\label{future}}
 \end{figure}

\bigskip
We can add a different kind of question: Who will solve these problems? We can hope that scientists of the future will not be limited by gender, ethnicity, or geographical location, as is already suggested by the increasing diversity of young people interested in science, like those of Fig.~\ref{future}.

Even though science has become increasingly collaborative, one of the major driving forces is still individual effort, for both theorists and experimentalists. And it is likely that the most creative scientists of the 21st Century will share characteristics with their famous predecessors, including a disdain for convention and compromise. 
Perhaps the greatest of scientists in antiquity was Archimedes, whose life was utterly different from ours. He had relatively meager resources outside his own mind. According to legend, his all-important geometric diagrams were drawn in sand, in ashes, and in oil on his body. He communicated in letters because there were no journals, and he did not have access to the convenient mathematical notations introduced almost two millennia later. His travels were relatively limited and difficult, as was daily living. 

Yet his approaches to theory, experiment, and invention were similar to those that we now take for granted.  We may have inherited these methods partly because he was a role model for the greatest minds of much later times, including Leonardo da Vinci, Galileo Galilei, and Isaac Newton. If Archimedes were alive today, he would surely appreciate our current standards for mathematical and scientific rigor, which are essentially the same as those of him and his contemporaries. The calculus he would certainly understand, since he came close to inventing it himself.

Archimedes was, of course, embedded in a scientific tradition that extends back to Thales of Miletus (who is said to have given the first recorded mathematical proof). And he had a highly individualistic (or eccentric) personal style, with a disregard for dress that achieves its limit when he is said to have run naked through the streets shouting ``Eureka!" after a major discovery. Newton and Einstein also paid little attention to personal attire or appearances, perhaps following Archimedes, but never reaching this extreme.

So, if Archimedes is a valid precedent, we can expect major breakthroughs from those who embed themselves in the current mainstream of scientific tradition, but who at the same time defy conventional points of view. 

\section*{Acknowledgements}

We are indebted to Stephen Fulling, Barry Garraway, Ulrich Heinz, Vitaly Kocharovsky, and Vladislav Yakovlev for their helpful comments and suggestions.


\bigskip\bigskip

\begin{thebibliography}{999}

\bibitem{endsci}John Horgan, \textit{The End of Science: Facing the Limits of Knowledge in the Twilight of the Scientific Age} (Basic Books, 2015).

\bibitem{endphys}David Lindley, \textit{The End of Physics} (Basic Books, 1993).

\bibitem{curie-quote} Marie Curie, \textit{Pierre Curie: With Autobiographical Notes by Marie Curie}, translated by Charlotte and Vernon Kellogg (Macmillan, New York, 1923). 

\bibitem{ginzburg} V. L. Ginzburg, ``What problems of physics and astrophysics seem now to be especially important and interesting (thirty years later, already on the verge of XXI century)?'', Physics -- Uspekhi 42, 353 (1999).

\bibitem{duff-list} M. J. Duff, ``Top Ten problems in fundamental physics'', Int. J. Mod. Phys. A16, 1012 (2001).

\bibitem{carroll}Sean Carroll, \textit{The Big Picture: On the Origins of Life, Meaning, and the Universe Itself} (Dutton, 2016).

\bibitem{hilbert}Benjamin Yandell, \textit{The Honors Class: Hilbert's Problems and Their Solvers} (A.K. Peters, 2001).

\bibitem{weinberg-1989} S. Weinberg, ``The cosmological constant problem'', Rev. Mod. Phys. 61, 1 (1989).

\bibitem{weinberg-1987} S. Weinberg, ``Anthropic Bound on the Cosmological Constant'', Phys. Rev. Lett. 59, 2607 (1987).

\bibitem{riess} A. G. Riess et al., ``Observational evidence from supernovae for an accelerating universe and a cosmological constant'', Astron. J. 116, 1009 (1998).

\bibitem{perlmutter} S. Perlmutter et al., ``Measurement of $\Omega$ and $\Lambda$ from 42 high-redshift supernovae'', Astrophys. J. 517, 565 (1999).

\bibitem{acc-review-2013} D. H. Weinberg, M. J. Mortonson, D. J. Eisenstein, 
C. Hirata, A. G. Riess, and E. Rozog, ``Observational probes of cosmic acceleration'', Physics Reports 530, 87 (2013), arXiv:1201.2434 [astro-ph.CO].

\bibitem{fulling} We thank Stephen Fulling for emphasizing this point.

\bibitem{rovelli} C. Rovelli, ``Notes for a brief history of quantum gravity'', arXiv:gr-qc/0006061.

\bibitem{dewitt}B. S. DeWitt,``Quantum Theory of Gravity. I. The Canonical Theory''. Phys. Rev. 160, 1113 (1967).

\bibitem{ADM}R. Arnowitt, S. Deser, and C. W. Misner, ``The Dynamics of General Relativity'', in \textit{Gravitation: an Introduction to Current Research}, Louis Witten ed. (Wiley, 1962), 
arXiv:gr-qc/0405109.

\bibitem{witten}Michael Green, John Schwarz, and Edward Witten, \textit{Superstring Theory} (Cambridge University Press, 1988).

\bibitem{polchinski}Joseph Polchinski, \textit{String Theory} (Cambridge University Press, 2005).

\bibitem{beckers}Katrin Becker, Melanie Becker, and John Schwarz, \textit{String Theory and M-Theory: A Modern Introduction} (Cambridge University Press, 2007).

\bibitem{schwarz}J. H. Schwarz, ``Introduction to Superstring Theory'', arXiv:hep-ex/0008017.

\bibitem{greene}Brian Greene, \textit{The Elegant Universe: Superstrings, Hidden Dimensions, and the Quest for the Ultimate Theory} (Norton, 2010).

\bibitem{kaku}Michio Kaku and Jennifer Thompson, \textit{Beyond Einstein: The Cosmic Quest for the Theory of the Universe} (Anchor, 1995).

\bibitem{randall}Lisa Randall, \textit{Warped Passages: Unraveling the Mysteries of the Universe's Hidden Dimensions} (Harper Perennial, 2006).

\bibitem{thorne}Kip Thorne, \textit{Black Holes and Time Warps: Einstein's Outrageous Legacy} (Norton, 1995).

\bibitem{bets}\url{https://en.wikipedia.org/wiki/Thorne–Hawking–Preskill_bet}.

\bibitem{susskind}Leonard Susskind, \textit{The Black Hole War: My Battle with Stephen Hawking to Make the World Safe for Quantum Mechanics} (Back Bay Books, 2009).

\bibitem{Interstellar}Kip Thorne, \textit{The Physics of Interstellar} (Norton, 2014).

\bibitem{guth}A. H. Guth, ``The Inflationary Universe: A Possible Solution To The Horizon And Flatness Problems'', Phys. Rev. D 23, 347 (1981).

\bibitem{kolb}E. Kolb and M. Turner, \textit{The Early Universe} (Addison-Wesley, 1990).

\bibitem{linde}A. Linde, ``Inflationary Cosmology after Planck 2013'', arXiv:1402.0526 [hep-th].

\bibitem{planck}Planck Collaboration: P. A. R. Ade et al., ``Planck 2013 results. XVI. Cosmological parameters'', arXiv:1303.5076 [astro-ph.CO].

\bibitem{trodden}M. Trodden, ``Electroweak baryogenesis'', Rev. Mod. Phys. 71, 1463 (1999).

\bibitem{peter}A. H. G. Peter, ``Dark Matter'', arXiv:1201.3942 [astro-ph.CO].

\bibitem{ong}R. A. Ong, ``Astroparticle physics'', Phys. Scr. T158, 014022 (2013).

\bibitem{bergstrom}L. Bergstr\"{o}m, ``Cosmology and the dark matter frontier'', Phys. Scr. T158, 014014 (2013).

\bibitem{cline}D. B. Cline, ``The search for dark matter'', Phys. Scr. 91, 033008 (2016).

\bibitem{freese}Katherine Freese, \textit{The Cosmic Cocktail: Three Parts Dark Matter}  (Princeton University Press, 2014).

\bibitem{cheng}T.-P. Cheng and L.-F. Li, \textit{Gauge Theory of Elementary Particle Physics} (Clarendon, 1984).

\bibitem{schwartz}M. D. Schwartz, \textit{Quantum Field Theory and the Standard Model} (Cambridge University Press, 2014).

\bibitem{pdg-2012}K.A. Olive et al. (Particle Data Group), Chin. Phys. C 38, 090001 (2014), updated at http://pdg.lbl.gov/. 

\bibitem{chierici-sym}R. Chierici, ``Unveiling the top secrets with the Large Hadron Collider'', Phys. Scr. T158, 014007 (2013).

\bibitem{perez-sym}G. Perez, ``Top quark theory and the new physics searches frontier'', Phys. Scr. T158, 014008 (2013).

\bibitem{murayama}H. Murayama, ``Theory of Neutrino Masses and Mixings'', Int. J. Mod. Phys. A 17, 3403 (2002), arXiv:hep-ph/0201022.

\bibitem{conrad}J. M. Conrad, ``Neutrino experiments and the Large Hadron Collider: friends across 14 orders of magnitude'', Phys. Scr. T158, 014012 (2013).

\bibitem{parke}S. Parke, ``Neutrinos: theory and phenomenology'', Phys. Scr. T158, 014013 (2013).

\bibitem{ATLAS} ATLAS Collaboration, ``Observation of a new particle in the search for the Standard Model Higgs boson with the ATLAS detector at the LHC'', 	Phys. Lett. B716, 1 (2012), arXiv:1207.7214 [hep-ex].

\bibitem{CMS} CMS Collaboration, ``Observation of a new boson at a mass of 125 GeV with the CMS experiment at the LHC'', Phys. Lett. B716, 30 (2012), arXiv:1207.7235 [hep-ex].

\bibitem{read}A. L. Read, ``The discovery and measurements of a Higgs boson'', Phys. Scr. T158, 014009 (2013).

\bibitem{mariotti}C. Mariotti, ``Beyond standard model Higgs boson searches in the ATLAS and CMS experiments'', Phys. Scr. T158, 014010 (2013).

\bibitem{ellis}J. Ellis, ``Summary of the Nobel symposium on Large Hadron Collider results'', Phys. Scr. T158,  014020 (2013).

\bibitem{barbieri}R. Barbieri, ``Electroweak theory after the first Large Hadron Collider phase'', Phys. Scr. T158, 014006 (2013).

\bibitem{martin} S. P. Martin, ``A Supersymmetry Primer'', arXiv:hep-ph/9709356, and references therein.

\bibitem{parker} M. A. Parker, ``Beyond the Standard Model'', Phys. Scr. T158, 014015 (2013).

\bibitem{nima}N. Arkani-Hamed, ``Beyond the Standard Model theory'', Phys. Scr. T158 014023 (2013).

\bibitem{kounnas}  C. Kounnas, A. Masiero, D. V. Nanopoulos, and K. A. Olive, \textit{Grand Unification with and without Supersymmetry and Cosmological Implications} (World Scientific, 1984).

\bibitem{mohapatra} R. N. Mohapatra, \textit{Unification and Supersymmetry: 
The Frontiers of Quark-Lepton Physics} (Springer, 2010).

\bibitem{barger} V. D. Barger and R. J. N. Phillips, \textit{Collider Physics Updated Edition} (Addison-Wesley, 1997).

\bibitem{kostelecky}V. A. Kostelec\'{k}y and N. Russell, \textit{Data Tables for Lorentz and CPT Violation}, Rev. Mod. Phys. 83, 11 (2011), arXiv:0801.0287.

\bibitem{abdo}A. A. Abdo et al., ``A limit on the variation of the speed of light arising from quantum gravity effects'', Nature 462, 331 (2009).

\bibitem{altarelli}G. Altarelli, ``The Higgs: so simple yet so unnatural'', Phys. Scr. T158, 014011 (2013).

\bibitem{alekhin}S. Alekhin, A. Djouadi, and S. Moch, ``The top quark and Higgs boson masses and the stability of the electroweak vacuum", Phys. Lett. B 716, 214 (2012), arXiv:1207.0980 [hep-ph].

\bibitem{degrassi}G. Degrassi, S. Di Vita, J. Elias-Mir\'{o}, J. R. Espinosa, G. F. Giudice, G. Isidori, and A. Strumia, ``Higgs mass and vacuum stability in the Standard Model at NNLO'', JHEP 1208, 098 (2012), arXiv:1205.6497 [hep-ph].

\bibitem{quigg-clay}C. Quigg, ``Beyond Confinement'', arXiv:1301.4905 [hep-ph].

\bibitem{millennium}Keith Devlin, \textit{The Millennium Problems: The Seven Greatest Unsolved Mathematical Puzzles Of Our Time} (Basic Books, 2003).

\bibitem{deroeck}A. De Roeck, ``Experimental results on quantum chromodynamics: what is next?", Phys. Scr. 014001 (2013).

\bibitem{sjostrand}T. Sj\"{o}strand, ``Challenges for QCD theory: some personal reflections'', Phys. Scr. T158, 014002 (2013).

\bibitem{schukraft}J. Schukraft. ``Heavy ion physics at the Large Hadron Collider: what is new? What is next?", Phys. Scr. T158, 014003 (2013).

\bibitem{muller}Berndt M\"{u}ller, ``Investigation of hot QCD matter: theoretical aspects'', Phys. Scr. T158, 014004 (2013).

\bibitem{QCD-transition} Y. Aoki, G. Endr\"{o}di, Z. Fodor, S. D. Katz, and K. K. Szab\'{o}, ``The order of the quantum chromodynamics transition predicted by the standard model of particle physics'', Nature 443, 675 (2006).

\bibitem{Saskia}S. Mioduszewski, ``The quest to understand QCD matter using heavy nuclei in collisions'', Phys. Scr. 90, 108014 (2015).

\bibitem{Heinz} We wish to thank Ulrich Heinz for suggesting this generalization.

\bibitem{BCS} \textit{BCS: 50 Years}, edited by Leon Cooper and Dmitri Feldman (World Scientific, 2010).

\bibitem{kane} M. Z. Hasan and C. L. Kane, ``Topological insulators'', Rev. Mod. Phys. 82, 3045 (2010).

\bibitem{Kohn} W. Kohn, ``An essay on condensed matter physics in the twentieth century'', Rev. Mod. Phys. 71, S59 (1999).

\bibitem{zangwill} We thank Andrew Zangwill for pointing this out, with credit assigned to Sankar Das Sarma.

\bibitem{kaluza1}T. Appelquist, A. Chodos, and P.G.O. Freund, \textit{Modern Kaluza-Klein
Theories} (Addison-Wesley, Menlo Park, 1987).

\bibitem{kaluza2}\textit{Physics in Higher Dimensions}, Volume 2, edited by T. Piran and S. Weinberg (World Scientific, Singapore, 1986).

\bibitem{linde2}Andrei Linde, ``Prospects of Inflation'', Phys. Scr. T117, 40 (2005).

\bibitem{seiberg}N. Seiberg, ``Emergent Spacetime'', 	arXiv:hep-th/0601234.

\bibitem{brief}S. W. Hawking, \textit{A Brief History of Time} (Bantum, 1998), p. 190.

\bibitem{Krauss}Lawrence Krauss, \textit{A Universe from Nothing: Why There Is Something Rather than Nothing} (Atria Books, 2013)

\bibitem{aleksan}R. Aleksan, ``Future large scale accelerator projects for particle physics'', Phys. Scr. 014016 (2013).

\bibitem{cattai}A. Cattai, ``Trends in research and development for future detectors'', Phys. Scr. T158, 014017 (2013).

\bibitem{fortey}Richard Fortey, \textit{Life: An Unauthorised Biography -- A Natural History of the First Four Thousand Million Years of Life on Earth} (Flamingo, 1998).

\bibitem{universe}Jeffrey Bennett and Seth Shostak, \textit{Life in the Universe} (Pearson, 2016).

\bibitem{LUCA} M. C. Weiss, F. L. Sousa, N. Mrnjavac, S. Neukirchen, M. Roettger, S. Nelson-Sathi, and W. F. Martin, ``The physiology and habitat of the last universal common ancestor'', Nature Microbiology 1, 16116 (2016).

\bibitem{lane}Nick Lane, \textit{The Vital Question: Energy, Evolution, and the Origins of Complex Life} (Norton, 2016).

\bibitem{Human} Brian Cox and Andrew Cohen, \textit{Human Universe} (Collins, 2015).

\bibitem{road}Frank Wilczek, \textit{Fantastic Realities: 49 Mind Journeys And a Trip to Stockholm} (World Scientific, 2006).

\bibitem{CKM-prize}``The Nobel Prize in Physics 2008'', \url{http://www.nobelprize.org/nobel_prizes/physics/laureates/2008/}.

\bibitem{Nir-flavor}Y. Nir, ``Flavor physics: past, present, future'', Phys. Scr. T158, 014005 (2013).

\bibitem{Gibson-flavor}V. Gibson, ``Heavy flavour at the Large Hadron Collider'', Phys. Scr. T158, 014021 (2013).

\bibitem{Krizan-flavor}P. Kri\u{z}an, ``Flavour physics at B factories'', Phys. Scr. T158, 014024 (2013).

\end{thebibliography}
\end{document}